\documentclass[aps,11pt]{revtex4-2}

\usepackage{graphicx}

\usepackage{float}
\usepackage[caption=false]{subfig}

\usepackage{amsmath}

\usepackage{upgreek}

\newcommand{\me}{\mathbf{e}}
\newcommand{\mr}{\mathbf{r}}

\usepackage{tabularx}

\begin{document}

\title{\Large{Modal properties of dielectric bowtie cavities \\ with deep sub-wavelength confinement}}

\author{George Kountouris $^1$, Jesper M{\o}rk$^1$, Emil Denning$^{1,2}$ and Philip Tr{\o}st Kristensen$^1$ \\
\vspace{5mm}
\scriptsize{1. Technical University  of  Denmark, Department of Electrical and Photonics Engineering, DK-2800  Kgs.   Lyngby,  Denmark, \\ Center for Nanophotonics -- NanoPhoton, Technical University  of  Denmark, DK-2800  Kgs.   Lyngby,  Denmark} \\ \vspace{2mm}
\scriptsize{2. Technische Universit{\"a}t Berlin, Institut f{\"u}r Theoretische Physik, Nichtlineare \\
\vspace{-3mm} Optik und Quantenelektronik, Hardenbergstrasse 36, 10623 Berlin, Germany} \vspace{1mm}}

\begin{abstract}
We present a design for an optical dielectric bowtie cavity which features deep sub-wavelength confinement of light.  The cavity is derived via simplification of a complex geometry identified through inverse design by topology optimization, and it successfully retains the extreme properties of the original structure, including an effective mode volume of $V_\text{eff}= 0.083 \pm 0.001 \; (\lambda_\text{c}/2n_\text{Si})^3$ at its center.  Based on this design, we present a modal analysis to show that the Purcell factor can be well described by a single quasinormal mode in a wide bandwidth of interest. Owing to the small mode volume, moreover, the cavity exhibits a remarkable sensitivity to local shape deformations, which we show to be well described by perturbation theory. The intuitive simplification approach to inverse design geometries coupled with the quasinormal mode analysis demonstrated in this work provides a powerful modeling framework for the emerging field of dielectric cavities with deep sub-wavelength confinement.
\end{abstract}

%\keywords{bowtie cavity, subwavelength, optical resonator, light-matter interaction, photonic crystal, Purcell factor, silicon}

\maketitle

\vspace{-15mm}

\section{Introduction}
\vspace{-5mm}

Dielectric bowtie cavities (DBCs) represent an emerging class of optical resonators in which light is concentrated to length scales much smaller than the wavelength in the material \cite{Gondarenko2006, Lu2013, Hu2016, Choi2017, Wang2018}.  This regime of deep sub-wavelength confinement has interesting implications for light-matter interaction and for a wide range of applications in the broad area of nanophotonics. In contrast to plasmonic structures, which can be subject to high nonradiative losses~\cite{Wang2006}, dielectric cavities, e.g. made out of Silicon or III-V semiconductors, generally feature much lower energy dissipation and are thus of particular interest for many applications.  In general, light-matter interaction with resonant fields and point-like emitters in optical cavities can be characterized by the quality factor $Q$, related to the temporal confinement of the light, and the effective mode volume $V_\text{eff}$, related to the electric field strength at the position of the emitter.  In the weak coupling regime, the relative enhancement in the radiative decay rate of a dipole emitter is given by the Purcell factor \cite{Purcell1946}
\begin{equation}
F_\text{P} = \frac{3}{4\pi^2} \left(\frac{\lambda_\text{c}}{n}\right)^3 \frac{Q}{V_{\mathrm{eff}}},
\label{Eq:Purcell_formula}
\end{equation}

\noindent where $\lambda_\text{c}$ and $n$ denote the resonance wavelength and the refractive index, respectively. This expression assumes that the spectral width of the cavity is much larger than the spectral width of the emitter.  Photonic structures realizing high Purcell factors are of interest for single-photon sources \cite{Ryu2003, Singh2006, MangaRao2007, Claudon2010, Ding2016, Pelton2015, Bozhevolnyi2016, IlesSmith2017}, and may also enable nano-LEDs with high bandwidth and noise squeezing\cite{Suhr2010, Mork2020}.  Beyond the weak coupling regime, increasingly strong coupling can lead to other interesting effects, such as phonon decoupling for near-unity indistinguishability \cite{Denning2020}, non-Markovian effects \cite{BundgaardNielsen2021, shahbazyan2021nonmarkovian} and even ultra-strong coupling \cite{Fleming_2010}. Whereas there is a rich and varied literature on well-established optical cavity designs\cite{Vahala2003, Benson2006} such as ring resonators~\cite{Bogaerts2011, Sarkaleh2017, Sacher2008, Steglich2019}, whispering-gallery resonators~\cite{Armani2003, Knight1995, Gorodetsky1996,Cai2020, Jiang2020}, photonic crystal defects \cite{Akahane2003, Matsuo2010, Birowosuto2014, Zhang2015, Saldutti2021}, or micropillars~\cite{Gerard1996, Takemoto2015, Osterkryger2019, Song2019, Peinke2021}, research into DBCs is still relatively young. The use of inverse design algorithms~\cite{Gondarenko2006, Liang2013, Wang2018} has shown that the bowtie feature arises naturally when computationally maximizing field localization, and it is by now understood that DBCs work by exploiting the electromagnetic interface conditions across high contrast refractive index regions \cite{Almeida2004, Hu2016, Choi2017, Albrechtsen2022}. Indeed, the bowtie geometry has also been derived from analytical considerations~\cite{Hu2016, Choi2017}, in what can be considered a natural progression of the ideas of Robinson et al.~\cite{Robinson2005}. A related earlier concept is that of the slot waveguide \cite{Almeida2004, Xu2004, Robinson2005, Barrios2009}, and it is interesting to note that bowtie waveguide designs have been suggested in Refs.~\cite{Yue2018, Sakib2019}.  Experimental realizations of DBCs are found in \cite{Hu2018} and \cite{Albrechtsen2021}.

\smallskip
The merits and potential impact of DBCs can be illustrated by comparing them with photonic crystal defect cavities, for which the effective mode volumes are consistently calculated and reported to be on the order of $(\lambda_\text{c}/2n)^3$; the current record holder appears to be the $H0$ cavity in Ref.~\cite{Minkov2014} with a reported effective mode volume of $V_\text{eff}=1.84(\lambda_\text{c}/2n)^3$. For DBCs, in contrast, values lower than $0.1(\lambda_\text{c}/2n)^3$ have been predicted, most recently in Refs.~\cite{Choi2017, Hu2016, Wang2018}, which is consistent with experiments showing a single spot of light localized in the center of the bowtie with SNOM-limited resolution \cite{Albrechtsen2021, Albrechtsen2020}. As more cavity designs for extreme light confinement are developed, considerations regarding their modal properties arise, such as a single-mode dominance, which is typically desirable for lasers, LEDs, and single-photon sources.  Additionally, mode analysis can provide convenient figures of merit depending on the application, usually via their $Q$-factors and some interaction mode volumes \cite{Notomi2010}.  Structures with highly localized fields present us with questions about sensitivity to disorder and fabrication processes, which can be expected to be especially relevant in DBC cavities due to their special confinement mechanism and delicate structures. To address these and related questions, in this work we present a thorough investigation of the modal properties of a generic DBC using the theoretical framework of quasinormal modes (QNMs)~\cite{Ching1998, Kristensen2020, Both2021}.

\medskip

To enable an effective and representative study of DBC cavities, we first present the use of an intuitive simplification approach by which we extract the main governing features of the design in Ref.~\cite{Albrechtsen2021}. Next, we show how one can use a QNM analysis to describe the electromagnetic response of the system with just a single QNM in a wide bandwidth of interest. Using this single QNM, we successfully show that the numerically optimized design in Ref.~\cite{Albrechtsen2021}, featuring a complex structure containing multiple spatial scales, can be replaced by a much simpler structure with practically no compromise in performance, within calculation error.  Finally, we use perturbation theory \cite{Lai1990,Johnson2002} to gauge the effect of shape deformations on the resonance and loss rate of the cavity, and we find that DBCs behave qualitatively differently from conventional photonic crystal cavities.  Most strikingly, the resonant wavelength shows a remarkable sensitivity, moving by about $19$ cavity linewidths when subject to a $1$-nanometer shift of the boundary.  Such variability in the bowtie size is typical in fabrication from e-beam exposure \cite{Albrechtsen2021}, and this sensitivity could be both a great asset and a challenge for future applications.

\medskip

This Article is organized as follows: In Section 2, we describe the simplifying design process and motivate it by discussing the use of topology optimization designs in practical calculations.  In Section 3, we define the Purcell factor, the Green tensor, and the QNMs, and subsequently show the Purcell spectrum of the cavity, demonstrating that a single mode approximation works extremely well in the vicinity of this resonance, and that the local response is dominated by this single mode over a very wide part of the spectrum.  In Section 4, we apply perturbation theory to investigate the effect of shifting the bowtie boundaries on the complex resonance of the mode, and demonstrate that it shows a remarkable sensitivity to local perturbations.  Finally, Section~\ref{Sec:Conclusion} holds the conclusions.

\vspace{-3mm}

\section{Topology optimization-inspired design}
\vspace{-3mm}

\begin{figure}
  \centering
    \subfloat[\label{fig1a}]{\includegraphics[width=0.30\columnwidth]{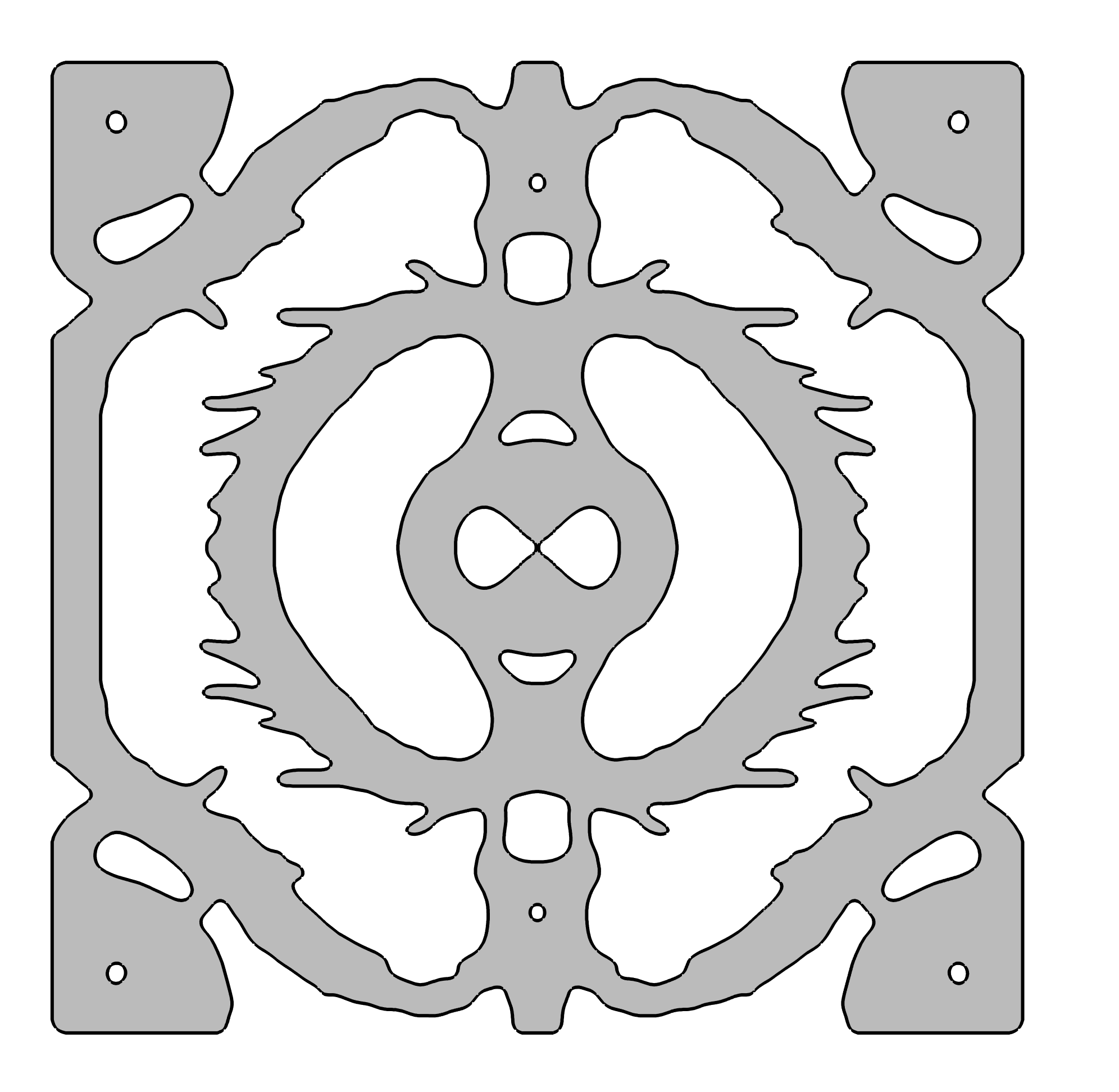}}
    \subfloat[\label{fig1b}]{\includegraphics[width=0.30\columnwidth]{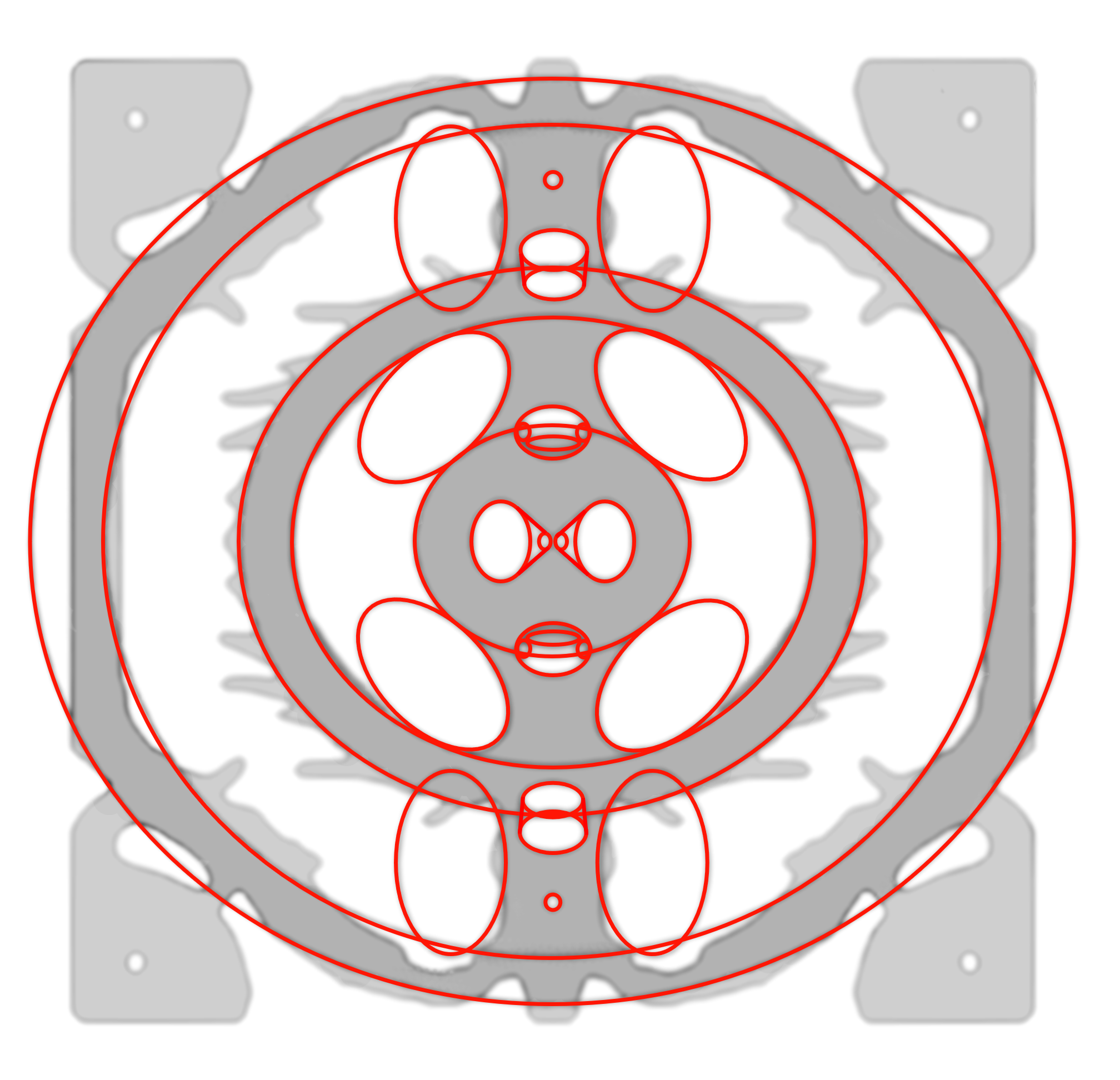}}
	\subfloat[\label{fig1c}]{\includegraphics[width=0.375\columnwidth]{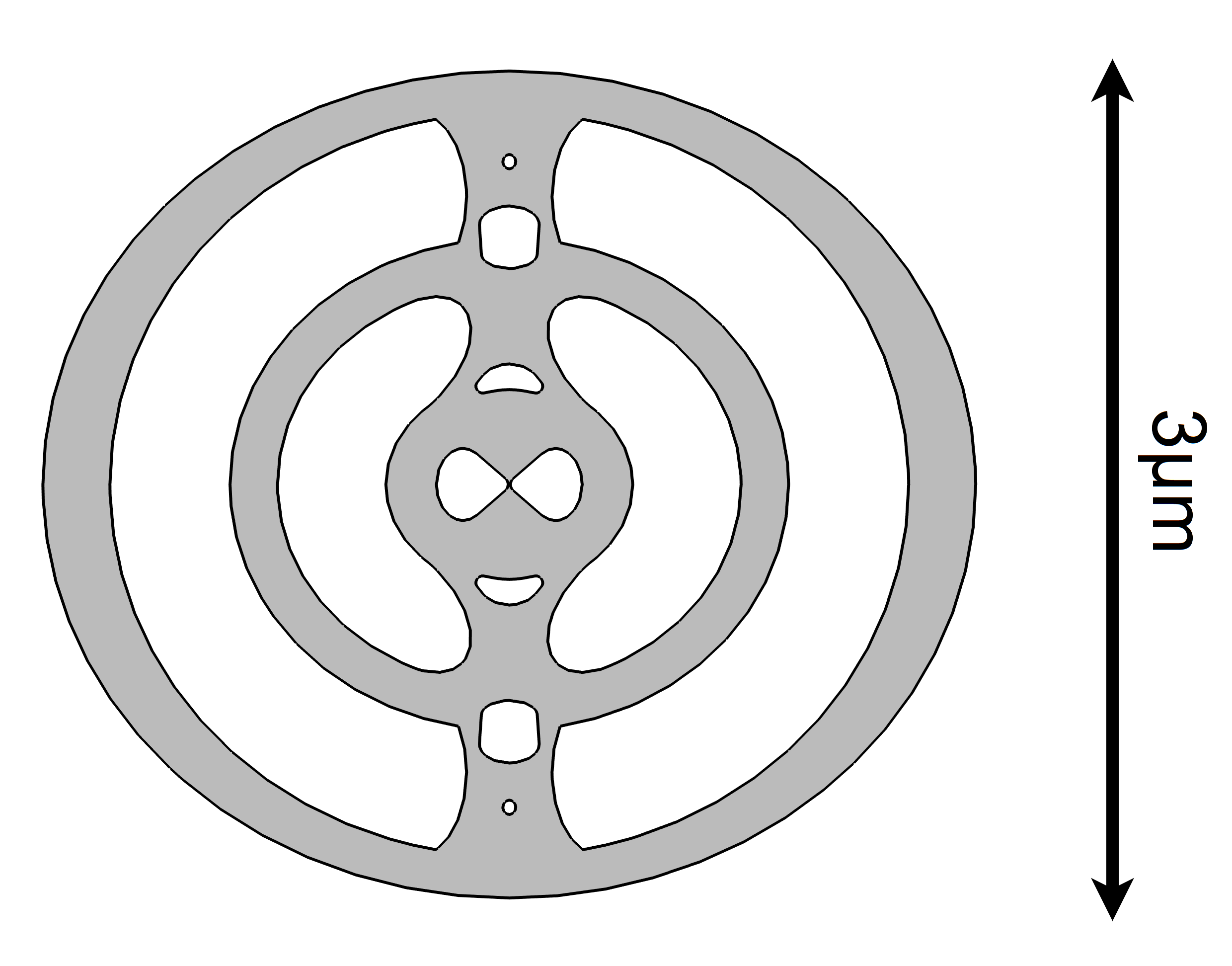}}
  \caption{From the TO design of Ref.~\cite{Albrechtsen2021} \textbf{(a)} to the extracted simplified cavity analyzed in this work \textbf{(c)}.  The raw TO design has sets of points defining its various air domains, visible in the first diagram.  The middle figure shows the geometrical simplifications overlaid on top of the original design.}
  \label{fig1}
\end{figure}

The geometry used in this work is inspired by the layout of the DBC in Ref. \cite{Albrechtsen2021} (cf. Fig.~\ref{fig1a}) which was created by the inverse design framework of topology optimization (TO) \cite{Christiansen2021} to maximize the local density of states inside the material in the cavity center, subject to constraints on the minimum feature sizes~\cite{Albrechtsen2021}. The TO method is a density-based approach to optimize the material distribution in a given design space by efficiently calculating gradients using adjoint sensitivity analysis \cite{Bendse1988, Bendse2004}.  In this way, the algorithm can provide locally optimized results of a specific figure of merit subject to given constraints \cite{Christiansen2021}.  For practical design and modeling purposes, however, it can be very useful to have an intuitive, generic design with smooth surfaces, where a few parameters can quickly and easily be adjusted, similar to what is often done by moving and modifying holes in photonic crystal defect cavities \cite{Akahane2003}. 

One practical challenge in directly using TO designs in modeling arises from the fact that the geometry is defined by a set of points. This is particularly relevant in finite element modeling, as these points typically define rigid nodes in the discretization and result in overmeshing or unnatural scaling of the mesh which, in practice, dramatically limits the feasibility of systematic mesh convergence studies as in this work. These meshing challenges have been addressed by various approaches to smoothen the TO designs in a partly or fully automated manner, as presented in Refs.~\cite{Miao2020, Liu2018, Swierstra2020}.  Here, we take a different approach and attempt to approximate the general shape of the TO design using only ellipses and tangents to realize a smooth final design.  In essence, we assume that the fine features in the TO design arise because of the requirement of optimization within a fixed rectangular domain, and their presence is not necessarily the only way to achieve the same results in the absence of a constrained optimization region.  Wang et al.~\cite{Wang2018} shows the influence of the calculation domain size on the emerging underlying structure, which is most clearly visible in regions further from the domain terminations, especially in the larger designs shown in Ref.~\cite{Wang2018}. The same tendency is visible in the design of Ref.~\cite{Gondarenko2006}.

The TO-inspired, simplified structure is shown along with the original TO design in Fig.~\ref{fig1}. In the middle diagram, the various simplifications made to the original are shown.  The design spans roughly $3\,\mathrm{\upmu m} \times 3\,\mathrm{\upmu m}$, with a thickness of $240\,\mathrm{nm}$ and a bowtie bridge width of $8\,\mathrm{nm}$ as in Ref.~\cite{Albrechtsen2021}. As we detail below, the fundamental QNM of interest in the simplified design has nearly the same mode volume and $Q$-factor as the original design, which corroborates the idea of the simplified design as an approximation to a primary underlying geometry.

\section{Electromagnetic response and modal analysis}

\begin{figure}
  \centering
    \subfloat{\includegraphics[width=0.55\columnwidth]{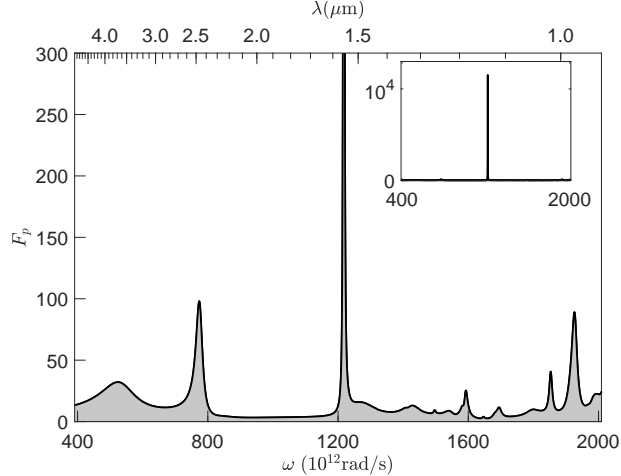}}
  \caption{Spectral dependence of the Purcell factor evaluated at a position $5\,\mathrm{nm}$ above the cavity center; note that the figure is cropped at $F_\text{P}=300$. The effect of the QNM of interest is visible as the dominant spike in the middle of the spectrum, which reaches a maximum on the order of ten thousand, as seen in the inset. }
  \label{fig2}
\end{figure}

As a convenient measure of the light-confining capabilities of the DBC, we calculate the Purcell factor of a dipole emitter at the position $\mr_0$ as the ratio of the imaginary part of the electromagnetic Green tensor to that in a homogeneous background material of refractive index $n_\text{B}$~\cite{Novotny2006},
\begin{align}
F_\text{P} = \frac{\text{Im}\left\{\me_\text{p}^\text{T}\cdot\mathbf{G}(\mr_0,\mr_0,\omega)\cdot\me_\text{p}\right\}}{\text{Im}\left\{\me_\text{p}^\text{T}\cdot\mathbf{G_\text{B}}(\mr_0,\mr_0,\omega)\cdot\me_\text{p}\right\}} = \frac{6\pi\text{c}}{n_\text{B}\omega}\text{Im}\left\{\me_\text{p}^\text{T}\cdot\mathbf{G}(\mr_0,\mr_0,\omega)\cdot\me_\text{p}\right\},
\label{Eq:Purcell_factor_from_Im_G}
\end{align}
where $\me_\text{p}$ is a unit vector in the direction of the dipole moment, and $\omega$ and $\text{c}$ denote the angular frequency and the speed of light, respectively. The Green tensor $\mathbf{G}(\mr,\mr',\omega)$ in general describes the electromagnetic response at the point $\mr$ due to a harmonically varying current source with frequency $\omega$ at the point $\mr'$. It is defined as the solution to the equation
\begin{equation}
\nabla\times\nabla\times \mathbf{G(r,r',\omega)} - k_0\epsilon_\text{r}(\mathbf{r})\mathbf{G(r,r',\omega)}=\delta(\mathbf{r-r'}),
\label{Eq:Gdef}
\end{equation}
where $k_0=\omega/\text{c}$ denotes the wavenumber, and $\epsilon_\text{r}(\mathbf{r})$ describes the relative permittivity, along with a suitable radiation condition to ensure that light propagates away from the cavity at large distances. The background Green tensor $\mathbf{G}_\text{B}$ is the solution to Eq.~(\ref{Eq:Gdef}) with $\epsilon_\text{r}(\mathbf{r})=\epsilon_\text{B}=n_\text{B}^2$. In practice, we find the Purcell factor by placing a point source at $\mr_0$ and calculating the scattered field at the same point; see Appendix~\ref{AppendixA} for details of the calculations and the methodology.  By calculating the Purcell factor spectrum at the position $5 \,\mathrm{nm}$ above the surface at the center and oriented along the bridge, it is immediately apparent that a single peak dominates the response, as shown in Fig. \ref{fig2}. Note that the figure is cropped at $F_\text{P}=300$ to show the relatively weak spectral structure surrounding the central peak, the top of which is on the order of ten thousand as seen in the inset. This peak, as we will show in the following paragraphs, can be very well described by a single-QNM approximation to the Green tensor.

\begin{figure}
  \centering
    \subfloat[\label{fig3a}]{\includegraphics[width=0.39\columnwidth]{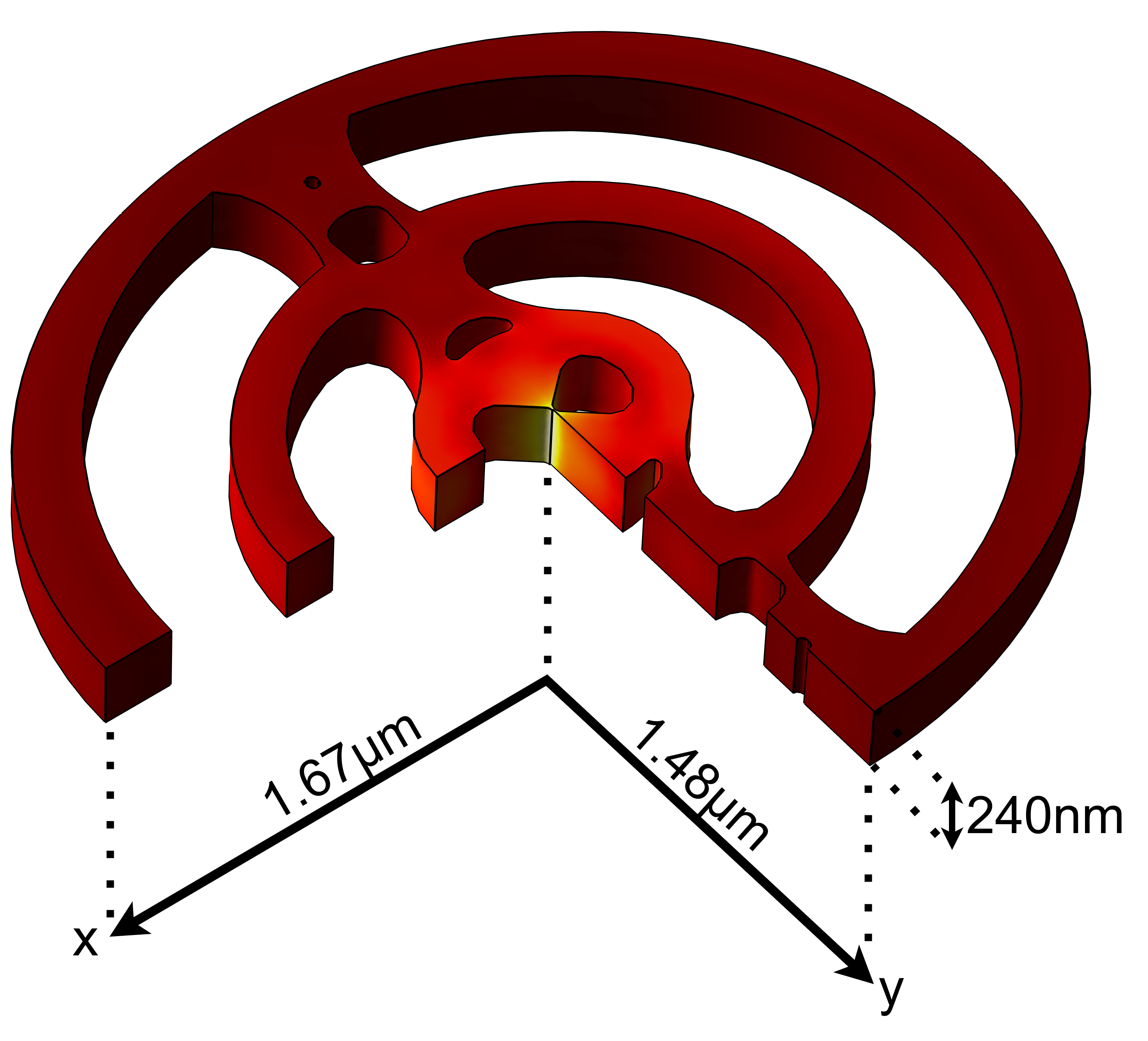}}
    \subfloat[\label{fig3b}]{\includegraphics[width=0.6\columnwidth]{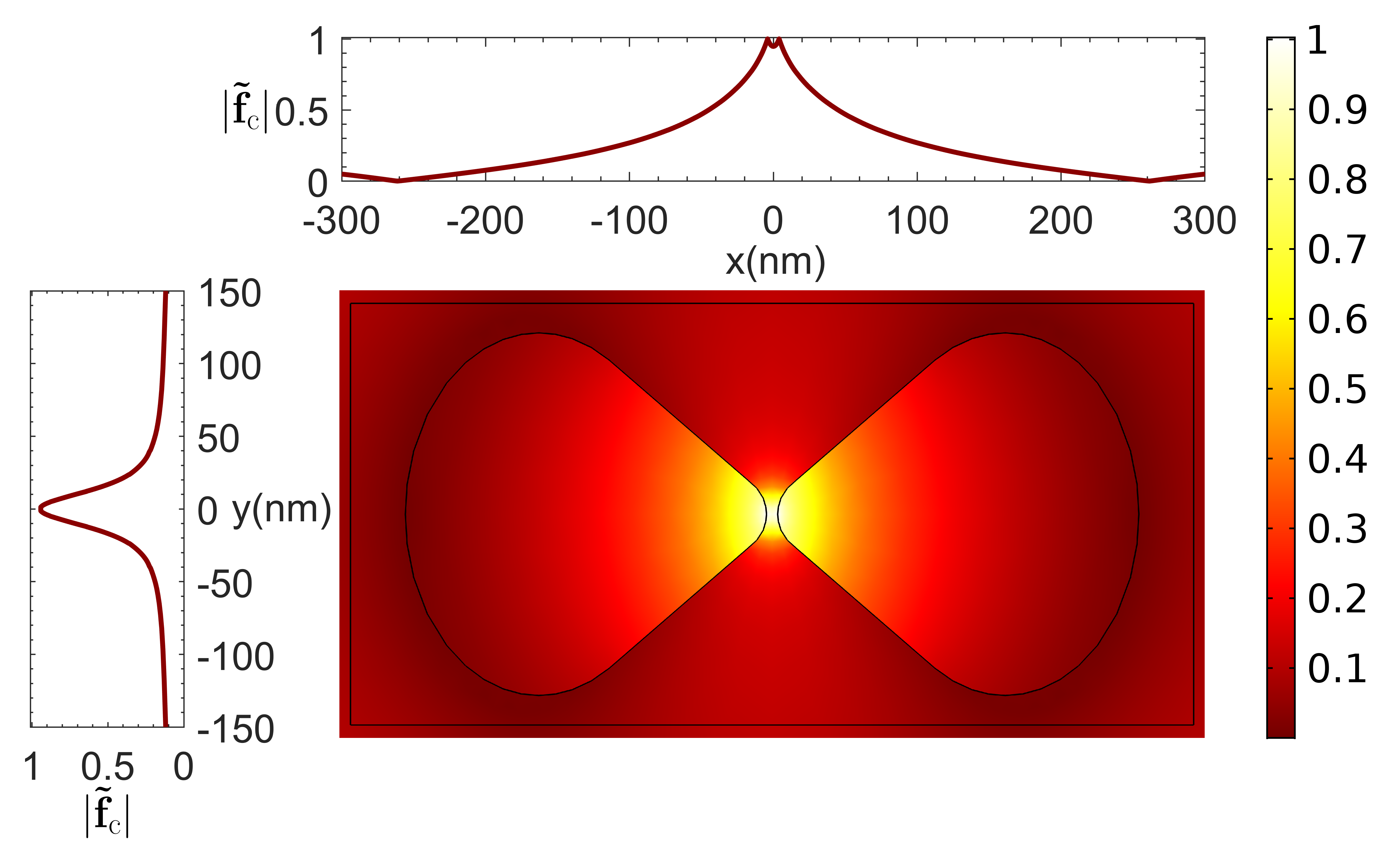}}
  \caption{\textbf{(a).} Schematic of the DBC (one quarter removed) showing the relative field strength $|\mathbf{\tilde{f}}_\text{c}|$ of the QNM of interest on the surfaces. \textbf{(b).} Zoom-in of the central bowtie structure. Top and left line graphs show the field strength along the horizontal and vertical lines through the center.}
  \label{fig3}
\end{figure}

The QNMs are solutions to the source-free electromagnetic wave equation subject to a suitable radiation condition~\cite{Kristensen2020}. For the electric field QNMs, we write the defining equation as
\begin{equation}
\nabla\times\nabla\times \tilde{\mathbf{f}}_\mu (\mathbf{r}) - \tilde{k}_\mu^2 %\left( \frac{\tilde{\omega}_\mu}{c} \right)^2
\epsilon_\text{r}(\mathbf{r})\tilde{\mathbf{f}}_\mu (\mathbf{r}) = 0,
\label{Eq:EwaveEq}
\end{equation}
where $\tilde{\mathbf{f}}_\mu (\mathbf{r})$ is the vectorial eigenmode, and we write the corresponding wavenumber as $ \tilde{k}_\mu=\tilde{\omega}_\mu/c$. For the calculations in this work, in which the cavity is surrounded by a homogeneous background material of refractive index $n_\text{B}$, we use the Silver-M{\"u}ller radiation condition,
\begin{equation}
\mathbf{\hat{r}}\times \nabla \times \tilde{\mathbf{f}}_\mu (\mathbf{r}) \rightarrow -\mathrm{i} n_{\mathrm{B}} \tilde{k}_\mu \tilde{\mathbf{f}}_\mu (\mathbf{r}),  \;\; r \rightarrow \infty,
\label{Eq:SMcondition}
\end{equation}
in which $\mathbf{\hat{r}}$ and $r$ denote, respectively, the direction and the magnitude of the radial position. Equations~(\ref{Eq:EwaveEq}) and (\ref{Eq:SMcondition})  define an eigenvalue problem, and as a result of the radiation condition, the solutions have complex eigenfrequencies $\tilde{\omega}_\mu = \omega_\mu - \mathrm{i}\gamma_\mu$, from which we can calculate the quality factor pertaining to each mode as $Q_\mu=\omega_\mu/2\gamma_\mu$. Figure \ref{fig3} shows the field profile of the mode of interest and its distribution within the material. In particular, it is clear that the field is very localized around the center of the bowtie, and drops quite quickly at positions further away. This is in accordance with the measurement results of \cite{Albrechtsen2020}. Measuring from the center of the cavity, the field magnitude $|\mathbf{\tilde{f}}_\text{c}(\mathbf{r})|$ drops to its half maximum at about $44 \,\mathrm{nm}$ along $x$, $16 \,\mathrm{nm}$ along $y$ and $125 \,\mathrm{nm}$ along $z$ ($5 \,\mathrm{nm}$ above the surface).

The magnetic field of the QNMs are related to the electric field as $\mathbf{\tilde{g}_\mu(r)} = -\text{i}\nabla\times\mathbf{\tilde{f}_\mu(r)}/\mu_0\tilde{\omega}_\mu$, and for ease of notation we combine them in a single entity as $\mathbf{\tilde{\underline{F}}_\mu(r)} = [ \mathbf{\tilde{f}_\mu(r)}, \mathbf{\tilde{g}_\mu(r)} ]^\text{T}$. The spatial field profiles of the QNM fields are divergent far from the resonator, which means that a special formulation is needed for their normalization. Several complementary formulations of the QNM normalization exist in the literature, as discussed in Ref.~\cite{Kristensen2015}; for this work we use the formulation~\cite{Muljarov2018, Kristensen2020}
\begin{align}
	\begin{split}
	\langle\langle  \mathbf{\tilde{\underline{F}}_\mu(r)} | \mathbf{\tilde{\underline{F}}_\mu(r)}  \rangle\rangle &= \frac{1}{2\epsilon_0}\int_V \left[ \epsilon_0\epsilon_\mathrm{r}(\mathbf{r})\mathbf{\tilde{f}_\mu(r)}\cdot \mathbf{\tilde{f}_\mu(r)} - \mu_0\mathbf{\tilde{g}_\mu(r)}\cdot \mathbf{\tilde{g}_\mu(r)}\right]\mathrm{dV}  \\
		&- \frac{\mathrm{i}}{2\epsilon_0\tilde{\omega}_\mu}\int_{\partial V} \left[ \left( r\partial_r\mathbf{\tilde{f}_\mu(r)}\right)\times\mathbf{\tilde{g}_\mu(r)} - \mathbf{\tilde{f}_\mu(r)}\times\left( r\partial_r\mathbf{\tilde{g}_\mu(r)}\right)\right]\cdot\hat{\mathbf{n}}\; \mathrm{dA}.
	\end{split}
	\label{norm_eqn}
	\end{align}
where $V$ is the volume of integration surrounding the cavity with boundary $\partial V$. Note that the dot product in the expression is without complex conjugation. Once normalized, the utility of a QNM description can be appreciated, if we express the Green tensor by use of a single-QNM approximation as \cite{Kristensen2020}
\begin{align}
\mathbf{G}(\mr,\mr',\omega) \approx \frac{\text{c}^2}{2\omega}\frac{\mathbf{\tilde{f}}_\text{c}(\mr)[\mathbf{\tilde{f}}_\text{c}(\mr')]^\text{T}}{\tilde{\omega}_\text{c}-\omega}.
\end{align}

\begin{figure}
  \centering
    \subfloat{\includegraphics[width=0.5\columnwidth]{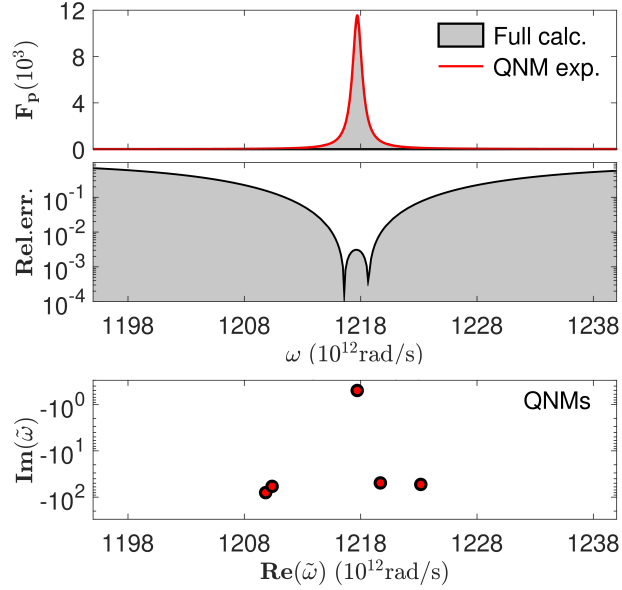}}
  \caption{Single QNM approximation of the Purcell factor $5 \,\mathrm{nm}$ above the surface of the cavity, and for a dipole moment along the $y$-direction. The top two figures show the single QNM expansion along with the relative error to the approximation, while the bottom figure shows the QNM spectrum in the same region.  The expansion uses the mode with the highest Q in the spectrum, which evidently provides the dominant contribution to the Purcell factor.}
  \label{fig4}
\end{figure}

By inserting in Eq.~(\ref{Eq:Purcell_factor_from_Im_G}), we get the single-QNM approximation to the Purcell factor. Figure \ref{fig4} shows the approximation to the Purcell factor at a position $5\,\mathrm{nm}$ above the surface along with a zoom-in of the reference calculation in Fig.~\ref{fig4}.  Also shown in Fig.~\ref{fig4} is the spectrum of complex QNM frequencies in the fourth quadrant of the complex plane. Even though several QNMs are seen to be present in the spectrum, they evidently contribute very little to the Purcell factor at the position of interest.  This is further illustrated by the central plot of the relative error when comparing to the reference calculation. Close to resonance, the relative error is as low as $0.3\%$. We emphasize that there are no fitting parameters as the two calculations are fully independent except for the fact that they were calculated using the same calculation mesh.

The approximately Lorentzian line-shape of the electromagnetic response is clearly visible in the top panel of Fig.~\ref{fig4}. On resonance at $\omega=\omega_\text{c}$, the expression reduces to the Purcell formula in Eq.~(\ref{Eq:Purcell_formula}), where now the effective mode volume is given as~\cite{Kristensen2012}
\begin{equation}
\frac{1}{V_{\mathrm{eff}}} = \mathrm{Re}\left\lbrace \frac{1}{v_\text{c}} \right\rbrace,
\end{equation}
in which 
\begin{align}
v_\text{c} = \frac{ \langle\langle \mathbf{\tilde{F}_\text{c}(\mr_0)} | \mathbf{\tilde{F}_\mu(r)}  \rangle\rangle }{\epsilon(\mathbf{r}_0) \mathbf{\tilde{f}}^2_\text{c}(\mathbf{r_0})}
\end{align}
is a generalized effective mode volume. From a convergence analysis, as detailed in Appendix~\ref{AppendixA}, we find the complex resonance frequency of the fundamental mode of interest to be $\omega_\text{c} = (1217.85 \pm 0.02) - \mathrm{i}(0.54\pm 0.02) \, 10^{12}\mathrm{rad \, s^{-1}}$. The corresponding real resonance wavelength is $1546.69 \pm 0.05 \,\mathrm{nm}$, and the $Q$-factor is $1222 \pm 28$. The on-resonance Purcell factor as calculated via Eq.~(\ref{Eq:Purcell_formula}) and for a dipole oriented along the bridge is $F_\text{P}=10592\pm 325$ at the position $5  \,\mathrm{nm}$ above the surface of the cavity center (as in Figs~\ref{fig2} and \ref{fig4}), and $F_\text{P}=8163\pm 250$ at the center of the cavity and hence inside the material. The corresponding effective mode volumes are $V_\text{eff}=0.064 \pm 0.001 \; (\lambda_\text{c}/2n_\text{air})^3$) and $V_\text{eff}=0.083 \pm 0.001 \; (\lambda_\text{c}/2n_\text{Si})^3$, respectively. For the original design, it was found in Ref.~\cite{Albrechtsen2021} that the resonance is at $1551 \,\mathrm{nm}$, with $Q\sim 1100$ and $V_{\mathrm{eff}}\sim0.08\;(\lambda_\text{c}/2n_\text{Si})^3$) in the cavity center.

\begin{figure}
  \centering
    \subfloat{\includegraphics[width=0.45\columnwidth]{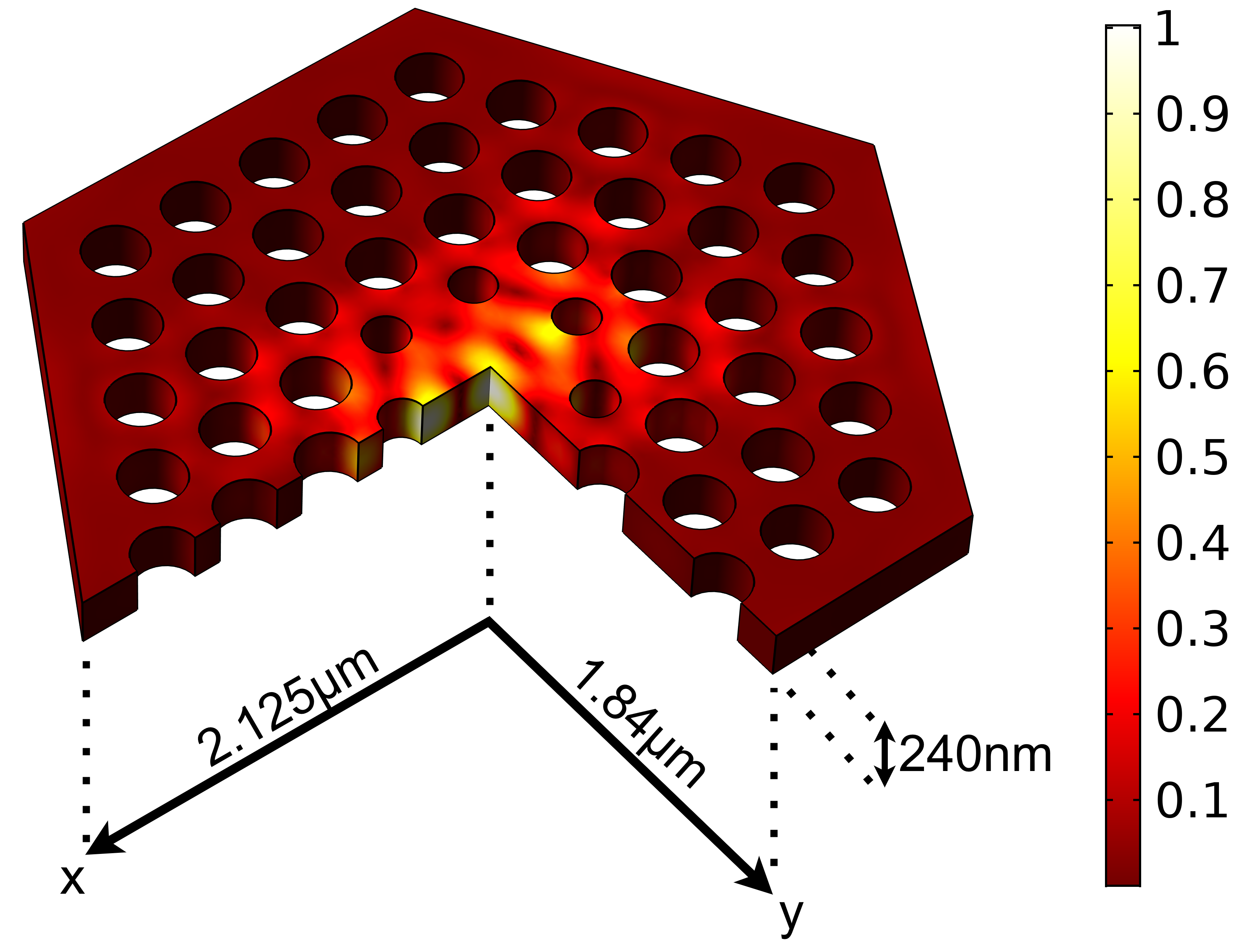}}
  \caption{Schematic of the $L1$ (one quarter removed) showing the relative field strength $|\mathbf{\tilde{f}}|$ of the QNM of interest on the surfaces.}
  \label{fig5}
\end{figure}

It is instructive to compare the DBC with a conventional photonic crystal cavity. To this end, we consider a so-called $L1$ cavity, where a single air-hole has been removed to act as a field-pinning defect in a membrane with hexagonal symmetry. The structure supports a fundamental QNM for which the electric field is shown in Fig.~\ref{fig5}.  

\begin{figure}
  \centering
    \subfloat[\label{fig6a}]{\includegraphics[width=0.45\columnwidth]{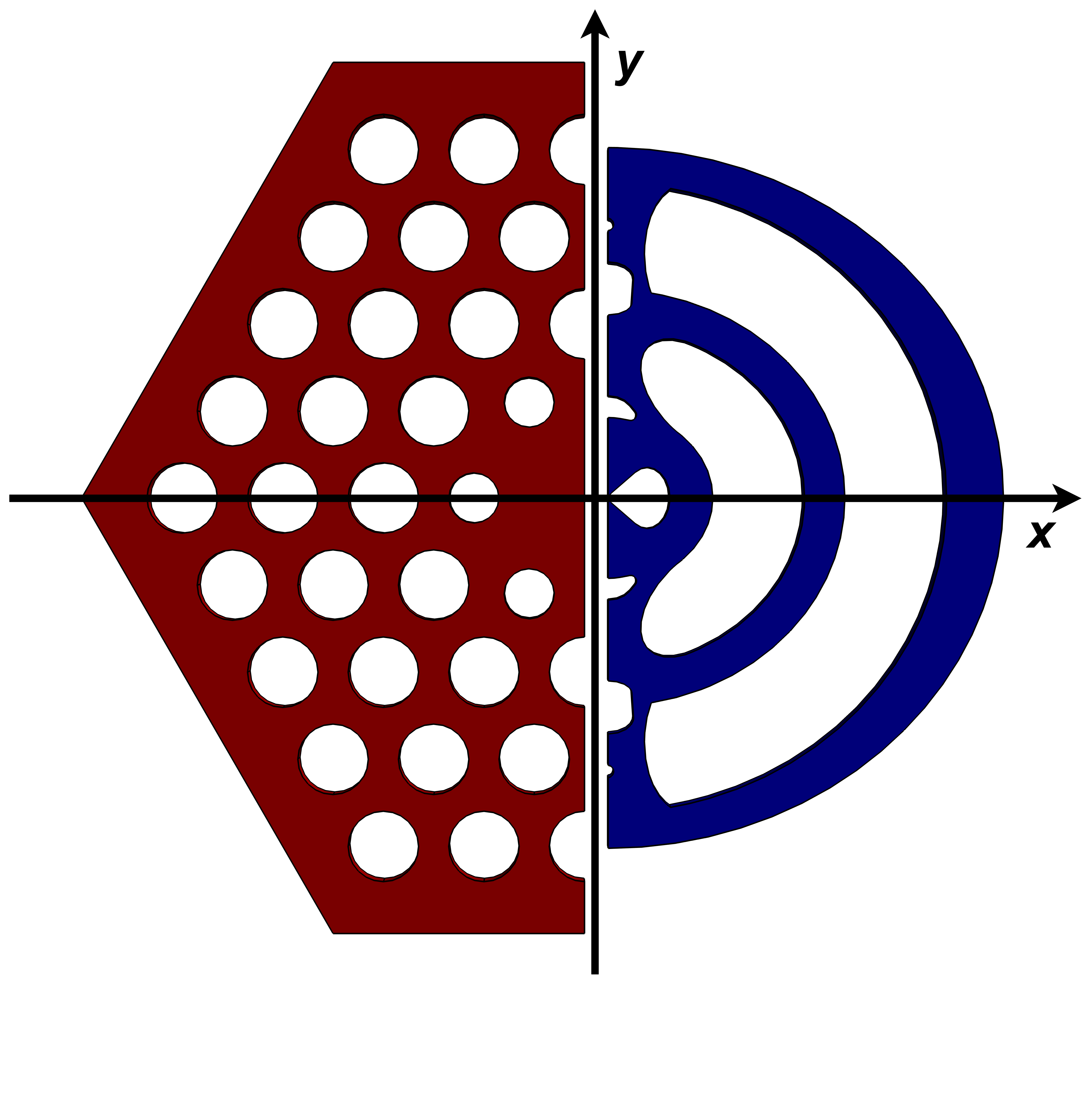}}
    \subfloat[\label{fig6b}]{\includegraphics[width=0.5\columnwidth]{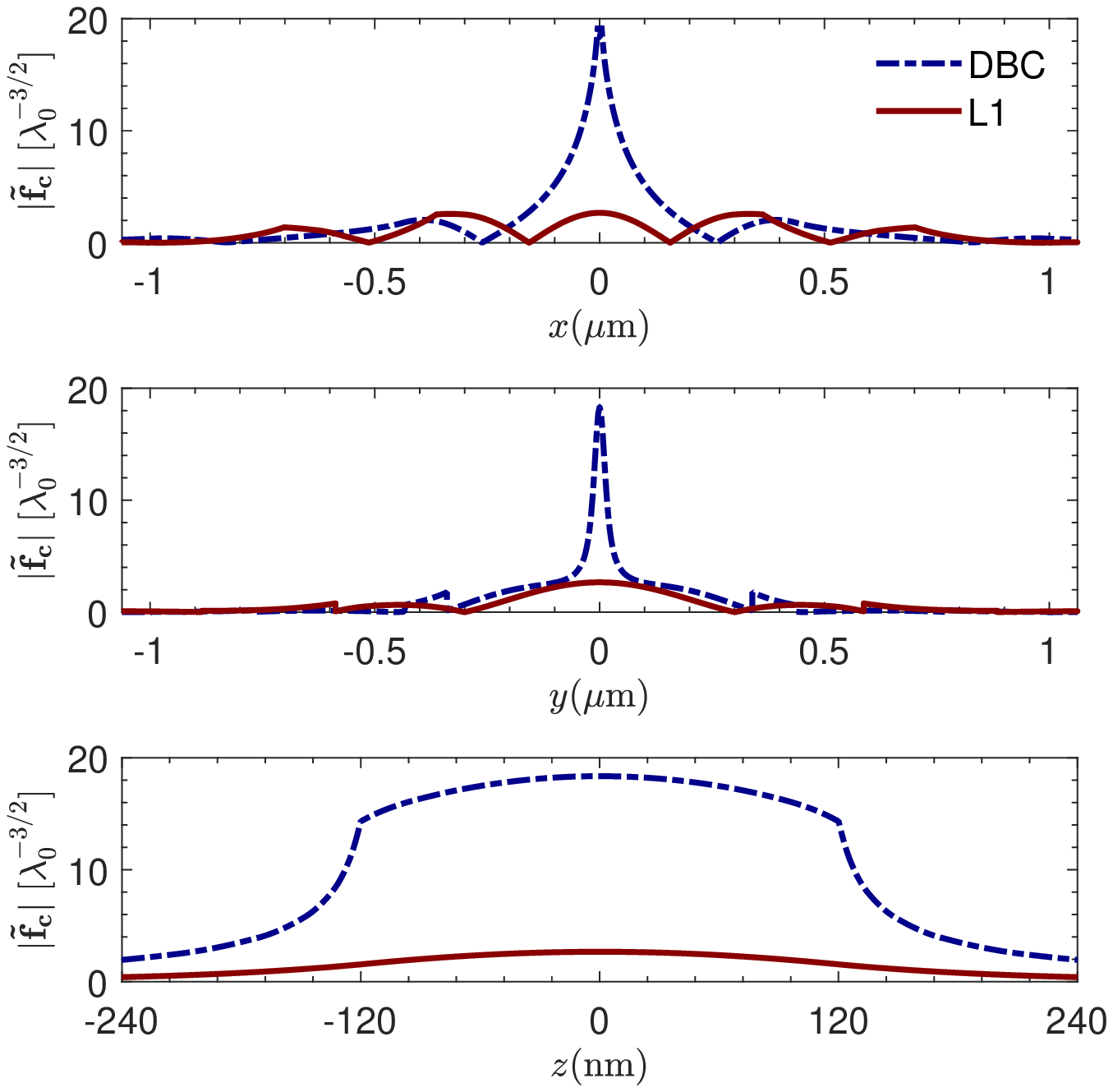}}
  \caption{\textbf{(a).} Diagram of the two cavities showing one half of each along with definitions of the axes. \textbf{(b).} A comparison of the field confinement between the $L1$ (solid red lines) and the DBC (dashed blue lines). The figures show the normalized mode profiles along the main axes $x$, $y$ and $z$, in units of $\lambda_0^{-3/2}$ with $\lambda_0=1550 \,\mathrm{nm}$.}
  \label{fig6} 
\end{figure}

The parameters for the reference cavity are based on the design in Ref.~\cite{Kim2012}, adjusted so that the resonance and quality factor are similar to those of the bowtie cavity of interest.  We use the same membrane thickness of $240 \,\mathrm{nm}$, a period $\alpha=425 \,\mathrm{nm}$, and a hole radius $R=0.35\alpha$, except for the holes closest to the cavity, which have been reduced in radius by $71\%$. We find the complex resonance frequency to be $\omega_\text{PC} = (1216.48 \pm 0.03) - \mathrm{i}(0.53\pm 0.03) \, 10^{12}\mathrm{rad \, s^{-1}}$. The corresponding real wavelength is $(1548.44\pm 0.04) \,\mathrm{nm}$, and the $Q$-factor is $Q=1162\pm 57$.  The effective mode volume at the center is $V_\text{eff}=3.9\pm 0.2 \; (\lambda_\text{PC}/2n_\text{Si})^3$), corresponding to a Purcell factor of $F_\text{P}=174\pm 9$, almost fifty times smaller than the bowtie cavity. Figure \ref{fig6b} compares the normalized field profiles of the two cavities along the primary axes as defined in Fig. \ref{fig6a}, showcasing the extreme confinement in the $xy$ plane of the DBC as compared to the L1 cavity.

\section{Geometrical perturbations}

\begin{figure}
  \centering
    \subfloat{\includegraphics[width=0.98\columnwidth]{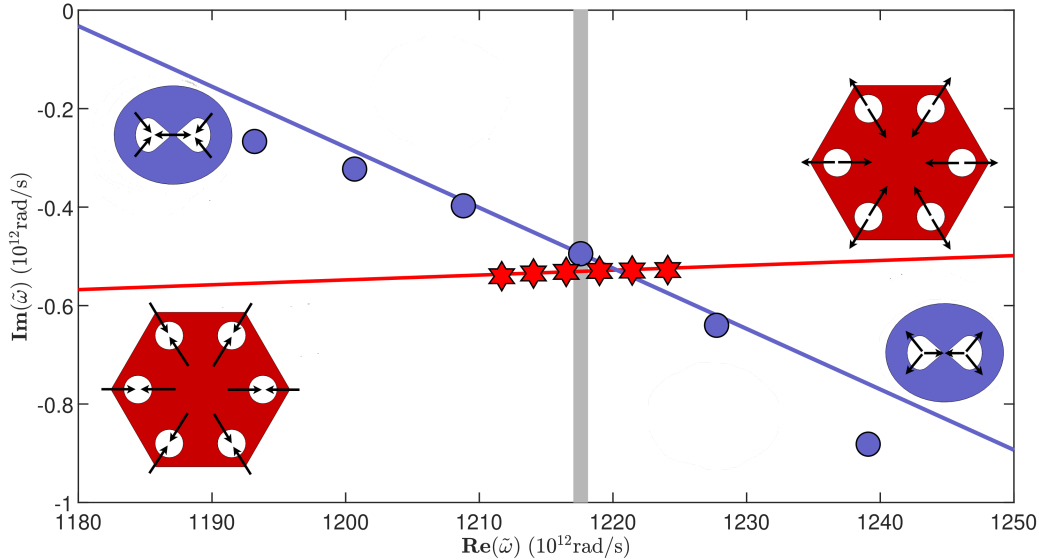}}
  \caption{Shape deformations of the bowtie and effect on the complex eigenfrequency, along with perturbation theory predictions (lines).  For comparison, also shown are the results for an L1 cavity under variation of its hole sizes.  The corresponding shifts are conceptually shown in the insets.}
  \label{fig7}
\end{figure}

Having established the validity of the single QNM approximation, we now use perturbation theory to analyze the effect of shape deformations. When considering shifting material boundaries, the first-order shift in the eigenfrequency $\Delta\tilde{\omega}_\mu$ is \cite{Lai1990,Johnson2002}
\begin{equation}
	\begin{split}
	\Delta\tilde{\omega}_\mu = &-\frac{\tilde{\omega}_\mu}{2}\int_{\partial V} \Bigl[ ( \epsilon_\text{R} - \epsilon_\text{B} )\mathbf{\tilde{f}^{||}_\mu(\mathbf{r})}\cdot\mathbf{\tilde{f}^{||}_\mu(\mathbf{r})} \\ 
	&- \left( \frac{\epsilon^2_\text{R/B}}{\epsilon_\text{R}} - \frac{\epsilon^2_\text{R/B}}{\epsilon_\text{B}} \right) \mathbf{\tilde{f}^{\perp}_\mu(\mathbf{r})}\cdot\mathbf{\tilde{f}^{\perp}_\mu(\mathbf{r})} \Bigr] \Delta h(\mathbf{r})  dA,
	\end{split}
\end{equation}
where $\epsilon_\text{R}$ is the relative permittivity of the resonator, placed in a background medium of relative permittivity $\epsilon_\text{B}$, and we consider a boundary shifted by $\Delta h(\textbf{r})$ along its normal direction.  The field normal $\mathbf{\tilde{f}^{\perp}_\mu(\mathbf{r})}$ and parallel $\mathbf{\tilde{f}^{\parallel}_\mu(\mathbf{r})}$ to the boundary is evaluated either on one or the other side of the boundary, reflected in the notation $\epsilon_\text{R/B}$, depending on the evaluation choice. We consider the case of shrinking or expanding the holes defining the bowtie, as illustrated in the insets of Fig. \ref{fig7}, for different perturbations $\Delta h = \{-2,-1,0,1,2,3\}  \,\mathrm{nm}$, with negative values denoting an enlargement of the bowtie gaps.  

To highlight the enhanced sensitivity to fabrication imperfections resulting from extreme confinement of the electromagnetic field in the DBC, we compare the results for the eigenfrequency shift and the perturbation theory predictions to those of the conventional $L1$ photonic crystal cavity in Fig.~\ref{fig5}, which has a similar footprint, resonance frequency, and $Q$, but for which the field confinement is much weaker. Figure \ref{fig7} shows the complex eigenfrequency shifts along with the predictions from perturbation theory.  The insets indicate the boundary shift direction in each case.  We find that the complex eigenfrequency shifts are markedly more pronounced in the case of the bowtie cavity, which is to be expected from QNM perturbation theory as shown by the straight lines in the plot. The vertical gray shading indicates the linewidth of the unperturbed structures. Owing to the light confinement, a shift of the sidewalls of just a single nanometer is enough to shift the resonance frequency of the DBC more than what is found for a three times larger shift for the PhC cavity, with $\Delta_\omega/\Delta_h = 9.5 \cdot 10^{12}\,\mathrm{rad \, s^{-1}nm^{-1}}$ against $\Delta_\omega/\Delta_h = 2.5 \cdot 10^{12}\,\mathrm{rad \, s^{-1}nm^{-1}}$, spanning several linewidths $\Delta\omega \approx 0.5 \cdot 10^{12}\,\mathrm{rad}\, s^{-1}$.  At the same time, and even more dramatically, the imaginary part changes at a rate of $0.12 \cdot 10^{12}\,\mathrm{rad \, s^{-1}nm^{-1}}$ for the bowtie cavity, while it is barely $0.0027 \cdot 10^{12}\mathrm{rad \, s^{-1}nm^{-1}}$ for the $L1$, about 44 times smaller. This quantifies how DBCs have resonances and $Q$-factors that are extremely sensitive to local perturbations compared to more conventional dielectric cavities. 

As a final note on the comparison, we remark that in the $L1$ cavity all of the holes in the structure are perturbed, whereas the analysis in Fig.~\ref{fig7} was performed by only perturbing the central bowtie of the DBC. To verify that this does not affect the conclusions, we performed similar calculations involving all the holes in the DBC (not shown) which resulted in less than $0.1\%$ difference in $\mathrm{Im}(\tilde{\omega}_\mu )$ and $\sim 0.7\%$ difference in $\mathrm{Re}(\tilde{\omega}_\mu )$ for the greatest perturbation considered, $\Delta h = -2  \,\mathrm{nm}$.  This further emphasizes the dramatic effect of changes to the bowtie, even for minute changes of a couple of nanometers. In practice, this effect places very tight constraints on the fabrication of optimized DBC structures, but successful implementations will benefit from enhanced light-matter interactions and may utilize the dramatic effect for sensing applications.

\section{Conclusion}
\label{Sec:Conclusion}

We have presented a small footprint, $3\,\mathrm{\upmu m} \times 3\,\mathrm{\upmu m}$, dielectric bowtie cavity with a deep sub-wavelength effective mode volume of $V_\text{eff}= 0.083 \pm 0.001 \; (\lambda_\text{c}/2n_\text{Si})^3$ at its center. The structure was derived by simplifying the topology optimization design in Ref.~\cite{Albrechtsen2021}, which was made to maximize the Purcell factor in the center of a finite domain.  Using this simplified design, we have shown that the Purcell factor can be very well approximated by use of a single quasinormal mode only, and we found the $Q$-factor and effective mode volume to be identical within calculation error to those found for the original and more complicated structure.  The simplicity of this so-called topology optimization-inspired design allows for much faster and more accurate finite element modeling, while simultaneously making it easier to manipulate for fundamental parameter studies, making the swift generation of alternative devices with different resonance frequencies or for other applications much more efficient.

\smallskip

As an illustration of the light-confining capabilities of dielectric bowtie cavities, we have compared the normalized mode profile of the fundamental quasinormal mode of interest to that of a more conventional $L1$ photonic crystal cavity with a similar resonance frequency and $Q$-factor, but a significantly larger effective mode volume. As an application of the modal analysis, moreover, we have used perturbation theory to analyze the effect of shape deformation on the central bowtie structure, showing greatly increased sensitivity of the complex quasinormal mode frequency compared to the reference cavity.

\smallskip

Dielectric bowtie cavities represent an emerging family of optical cavities with interesting properties, such as high sensitivity, strong confinement, low losses, and small footprints, which is important for fundamental research in light-matter interactions, and for applications ranging from quantum technology to new and improved lasers and detectors.  We believe the simplification approach to TO designs coupled with the quasinormal mode analysis shown in this work will provide a powerful modeling framework for studying this new class of cavities with sub-wavelength confinement.

\bigskip

%\begin{acknowledgement}
%The authors would like to thank Ole Sigmund and Rasmus Elleb{\ae}k Christiansen for developing and providing the original topology optimization design, as well as for fruitful discussions regarding topology optimization and optimizing the extracted simplified design.
%\end{acknowledgement}

%\begin{funding}
%This work was supported by the Danish National Research Foundation through NanoPhoton - Center for Nanophotonics, grant number DNRF147. Emil Vosmar Denning acknowledges support from Independent Research Fund Denmark through an International Postdoc Fellowship (Grant No. 0164-00014B).
%\end{funding}

\bibliographystyle{ieeetr}

\bibliography{Bibliography}

\newpage

\appendix
\section{Calculation methodology}
\label{AppendixA}
In this appendix, we give details of the calculation methods. For all calculations, we used the finite element method with curl conforming, curvilinear and isoparametric second-order tetrahedral elements, as implemented in COMSOL Multiphysics version 5.5.

\subsection{QNMs, Green tensor and boundary conditions}

In this section, we present details of the calculation of the QNMs and the Green tensor.  We also describe the implementation of the radiation condition and its limitations, which motivate in large part the use of the convergence studies described in the following section.

\bigskip

\begin{figure}
  \centering
    \subfloat[\label{figA1a}]{\includegraphics[width=0.49\columnwidth]{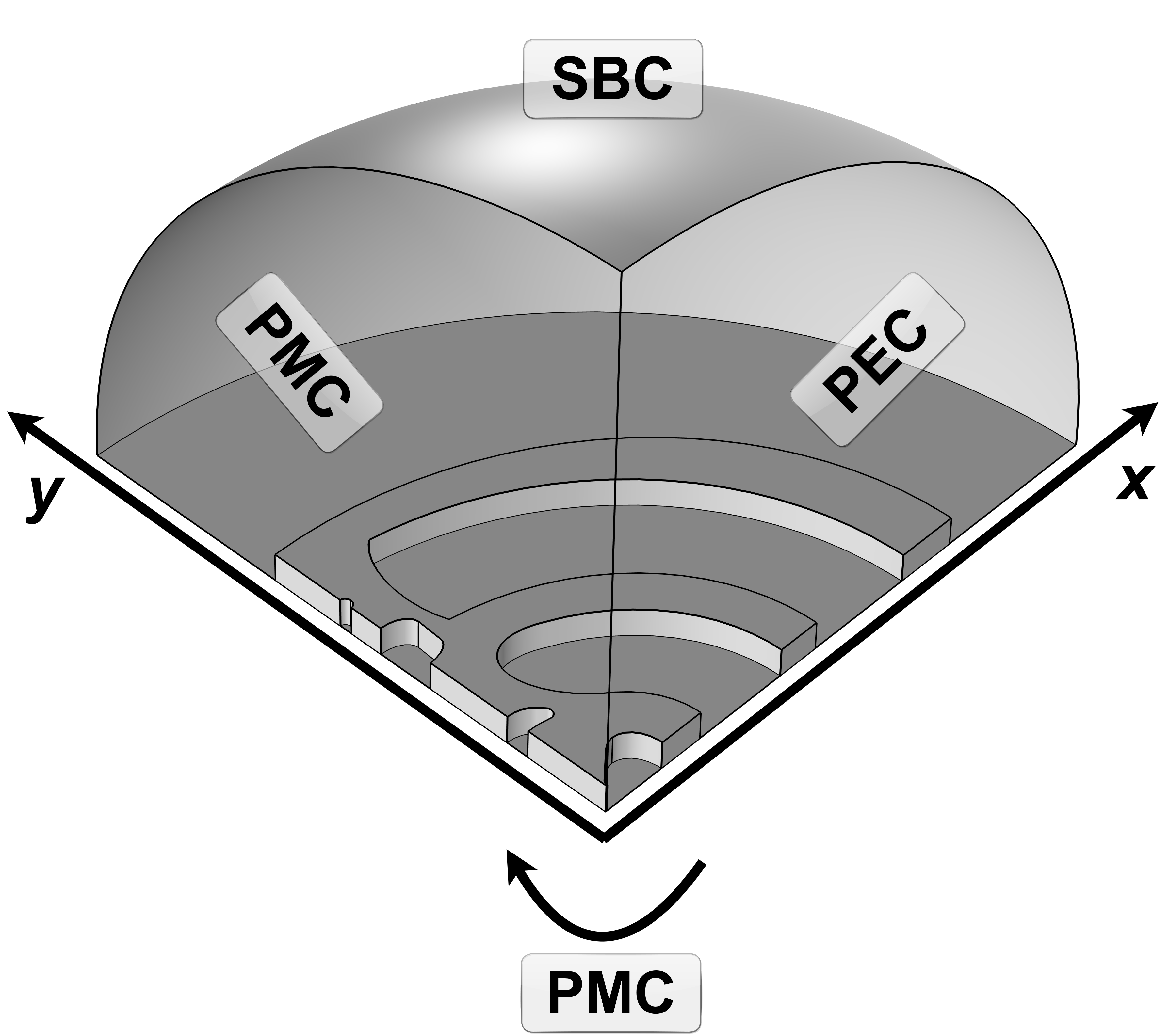}}
    \subfloat[\label{figA1b}]{\includegraphics[width=0.49\columnwidth]{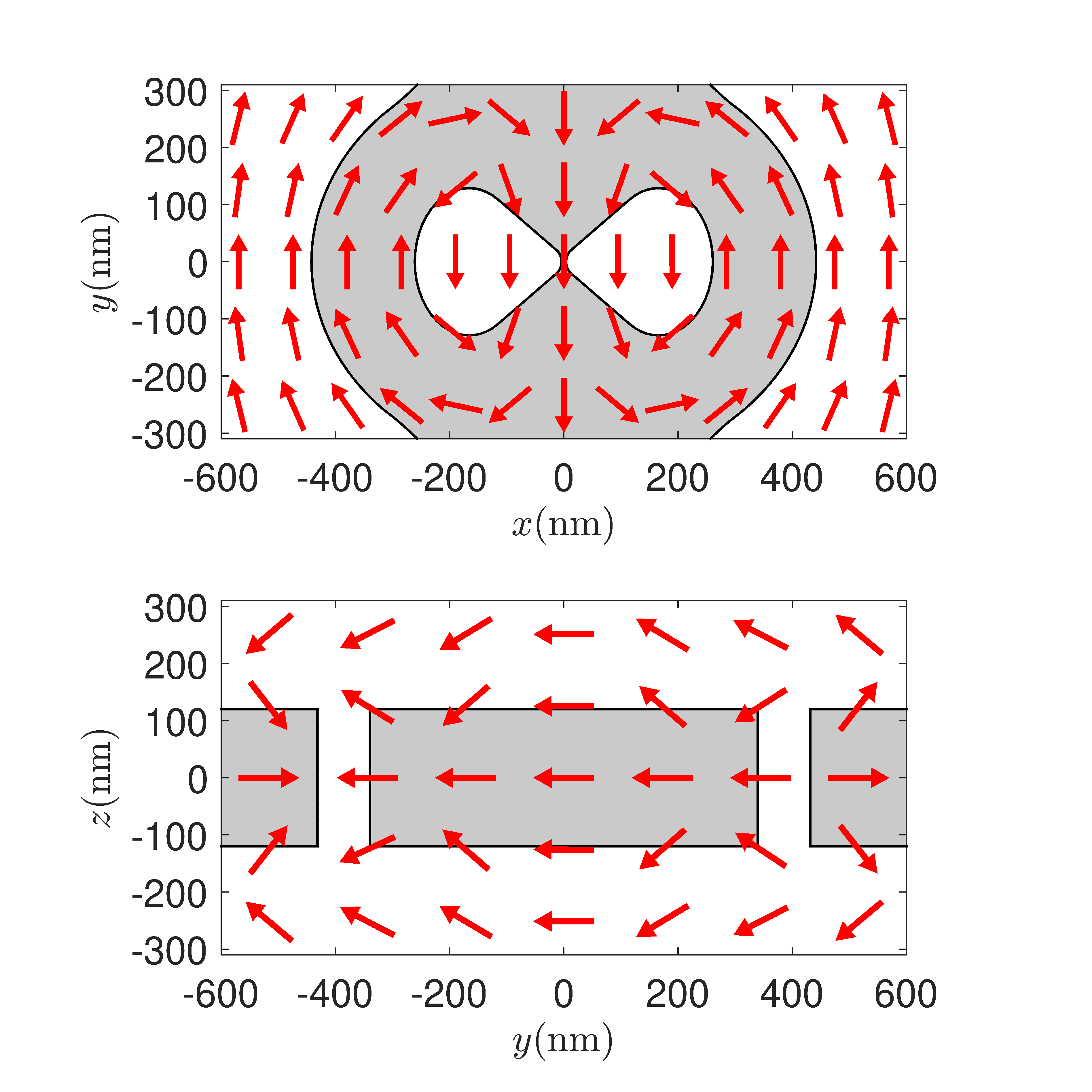}}
  \caption{\textbf{(a).} Symmetry-reduced model for the QNM convergence study.  PEC: perfect electric conductor, PMC: perfect magnetic conductor and SBC: scattering boundary condition. \textbf{(b).} Orientation of the electric-field components of the QNM on the $xy$ and $yz$ plane.}
  \label{figA1}
\end{figure}

For the convergence studies of the QNM of interest, we used a symmetry-reduced computational domain, as illustrated in Fig.~\ref{figA1}, in order to reduce the number of degrees of freedom to roughly an eighth compared to the original problem.  For a coordinate system as in the figure, the QNM of interest is symmetric with respect to the $xy$ and $yz$ planes, and antisymmetric with respect to $xz$.  Figure \ref{figA1b} shows arrow plots of the real part of the electric field QNM in the two planes, for which the symmetries in Fig. \ref{figA1a} are visible.

For the Green tensor calculations, we used a scattered-field formulation. To this end, we split the Green tensor in two parts corresponding to the background Green tensor and a scattered part as $\mathbf{G}(\mr,\mr',\omega) = \mathbf{G}_\text{B}(\mr,\mr',\omega)+\mathbf{G}_\text{S}(\mr,\mr',\omega)$, and write $\epsilon_\text{r} (\mathbf{r}) = \epsilon_\text{B}+\Delta\epsilon(\mathbf{r})$. Inserting in Eq.~(\ref{Eq:Gdef}) and rearranging the terms, we find that the scattered part of the Green tensor solves the equation
\begin{equation}
\nabla\times\nabla\times \mathbf{G}_\text{S}(\mr,\mr',\omega) - k_0^2 \epsilon_\text{r}(\mathbf{r})\mathbf{G}_\text{S}(\mr,\mr',\omega) = \frac{\omega^2}{c^2}\Delta\epsilon(\mathbf{r})\mathbf{G}_\text{B}(\mr,\mr',\omega).
\label{Eq:G_scat_def}
\end{equation}
In practice, for a given $\mr'$, each column of Eq.~(\ref{Eq:G_scat_def}) defines a wave equation for an electric field with a non-trivial source term defined by the background Green tensor. The background Green tensor is known analytically \cite{Novotny2006} and is known to diverge in the limit $\mr=\mr'$; this does not impact the calculations in this work, since we chose the point $\mr'$ to be in the free space background medium for which $\Delta\epsilon(\mathbf{r})=0$.

\smallskip

For the direct comparison of the Green tensor with the single QNM approximations in Fig.~\ref{fig4}, these calculations were performed using the exact same mesh and basis functions to make residual numerical errors in both calculations as similar as possible. Also, we explicitly inserted a node at the position of interest to avoid additional errors due to interpolation inside mesh elements of different sizes.

\medskip

Both the QNM and Green tensor problems require the application of an appropriate radiation condition.  We approximated the radiation condition in Eq.~(\ref{Eq:SMcondition}) --- which is defined in the limit $r\rightarrow\infty$ --- with the first order scattering boundary condition,
\begin{equation}
\mathbf{n}\times (\nabla\times \mathbf{E}(\mr))+\text{i}k \, \mathbf{n}\times(\mathbf{E}(\mr)\times\mathbf{n})=0,  %\;\; \mathbf{r} \in \partial V, %\rightarrow \infty
\label{Eq:first_order_scattering}
\end{equation}
in which $\mathbf{n}$ is the normal vector, and $\mathbf{E}(\mr)$ is an electric field. The wavenumber $k$ is real for the case of the Green tensor calculations, whereas it is complex for the QNM calculations. For the QNM calculations, we note that this boundary condition makes the eigenvalue problem non-linear.

\smallskip

We note that Eq.~(\ref{Eq:first_order_scattering}) is essentially the result of evaluating the radiation condition directly at the calculation domain boundary. For resonators with relatively high $Q$-factors, however, the condition works well also at finite distances, although residual reflections lead to small oscillations in the results as a function of calculation domain size, as shown in the following section.  Compounding the problem is that the mesh also has a major role in the calculation accuracy.  Nevertheless, it is possible to calculate high-accuracy values for key figures of merit from the calculations without needing exact boundary conditions, very large domain sizes, or extremely fine meshes, by use of convergence studies, as detailed below.

\subsection{Convergence studies}

\subsubsection{Numerical oscillations and discretization}
In this section we identify and discuss the main sources of error in the calculations and motivate the following convergence study.  There are mainly two parameters that determine the numerical error in the calculated eigenmodes. The first is the computational domain size, which is related to the scattering boundary condition, and the other is the discretization, which is related to the fineness and the order of the elements.  For practical reasons, we work with second-order elements and assess the influence of the mesh by considering the variations in the quantities of interest as we systematically decrease the characteristic length of the elements.

We start with an initial, relatively coarse mesh and refine the mesh by splitting the longest side of each element.  In our convergence studies, we use the average size of the elements as found through spatial averaging, i.e. $\overline{h} = \int h \rm{dV}/\int \rm{dV}$, where $h$ is the longest side of each element and the averaging is over some chosen volume.  We consider two volumes for the averaging; one is a spherical domain of radius $1.5\lambda_0, \lambda_0=1550 \,\mathrm{nm}$ and the other is the smallest, roughly cylindrical domain enclosing only the cavity region.  The resulting averages are denoted by $\overline{h}_\text{all}$ and $\overline{h}_\text{cav}$, respectively.  We found that the average size of the elements scales very systematically between most iterations. More specifically, subsequent iterations of this refinement process scale the average longest side of an element by a factor between $0.69-0.71$ for the last four iterations, which we use for the convergence study.  These results are summarized in Table ~\ref{Table1}.

\begin{table}  %[!htbp]
\centering
\begin{tabular}{ c c c c c c c  } 
Refinements & Elements    & Ratios        & $\overline{h}_\text{all}( \,\mathrm{nm})$ & 	Ratios	&  $\overline{h}_\text{cav}( \,\mathrm{nm})$ & Ratios \\ \hline
0           & 20683	      &  -	          &   298.01               		   & -		    & 97.610		                   & -      \\
1           & 74946	      & 3.6236	      &   230.78		               & 0.7744	    & 76.607		                   & 0.7848 \\
2           & 237995	  & 3.1756	      &   161.03		               & 0.6978		& 54.358		                   & 0.7096 \\
3           & 752303	  & 3.1610	      &   112.06		               & 0.6959		& 38.013		                   & 0.6993 \\
4           & 2136250	  & 2.8396	      &   79.628		               & 0.7106		& 26.813		                   & 0.7053 \\
\end{tabular}
\caption{Mesh-related parameters for the various refinements used in this study.  The first column indicates the number of refinements performed, while the second shows the corresponding number of elements.  $\overline{h}_\text{all}$ is the average element longest side in the entire domain of radius $1.5\lambda_0$, while $\overline{h}_\text{cav}$ is the average longest side in the roughly cylindrical region enclosing the cavity.  The ratios shown are between the finer and coarser iteration of the refinement process.}
\label{Table1}
\end{table}

\begin{figure}
  \centering
    \subfloat{\includegraphics[width=0.95\columnwidth]{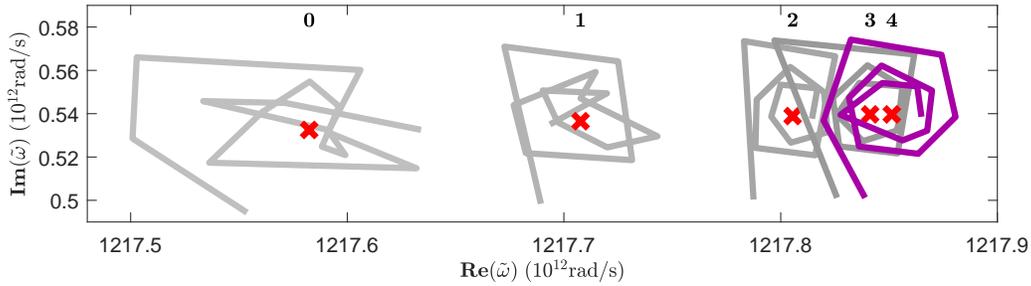}}
  \caption{Numerical oscillations of the eigenfrequency in the complex plane.  The data points show the complex eigenfrequencies calculated for expanding domain size with the same discretization.  The patterns appearing by connecting the data points spiral inwards, demonstrating the numerical convergence.  The leftmost spiral corresponds to the coarsest mesh, while the rightmost (purple) spiral corresponds to the fourth mesh refinement.  The red crosses show the extrapolated limiting values from each of these oscillations, as calculated with a moving average method.}
  \label{figA2}
\end{figure}

To illustrate the effect of varying both the discretization and the domain size, in Fig. ~\ref{figA2} we plot the complex eigenfrequency of the DBC in the complex plane for different discretizations and domain sizes.  We used spherical domains with radii in the range $1.5\lambda_0$ to $3\lambda_0$, with $\lambda_0=1550 \,\mathrm{nm}$, using steps of a tenth of $\lambda_0$ to adequately capture the numerical oscillations.  The discretization is indicated by a number from $0$ to $4$ corresponding to the number of refinements performed on the initial coarse mesh.  Each spiral comes from calculations with the same discretization and varying domain size, with the spirals closing inwards as the size increases.  We note that the spirals noticeably shift, and tend to become more regular with improving discretization.

\begin{figure}
  \centering
    \subfloat[\label{figA3a}]{\includegraphics[width=0.43\columnwidth]{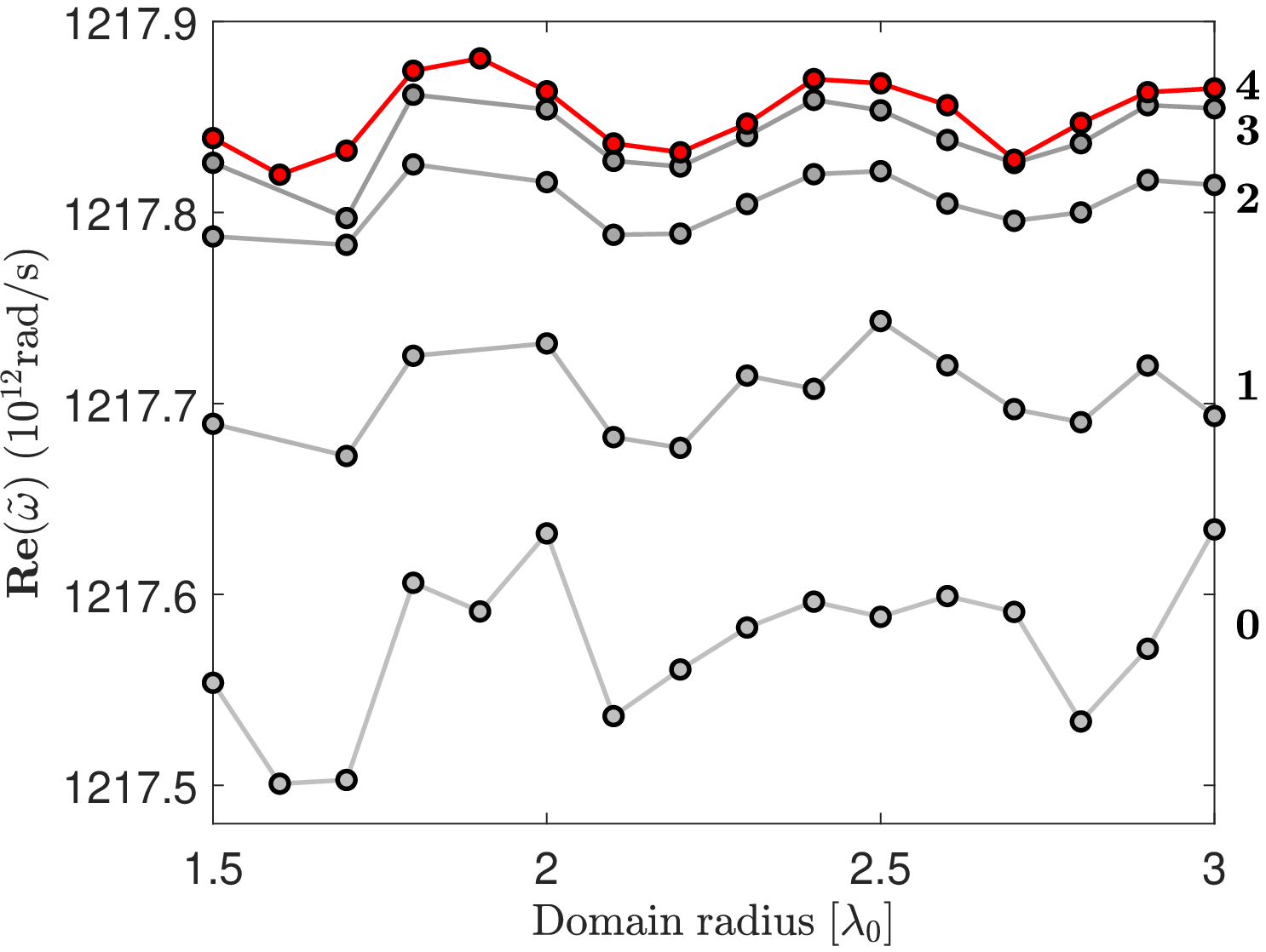}}
    \subfloat[\label{figA3b}]{\includegraphics[width=0.43\columnwidth]{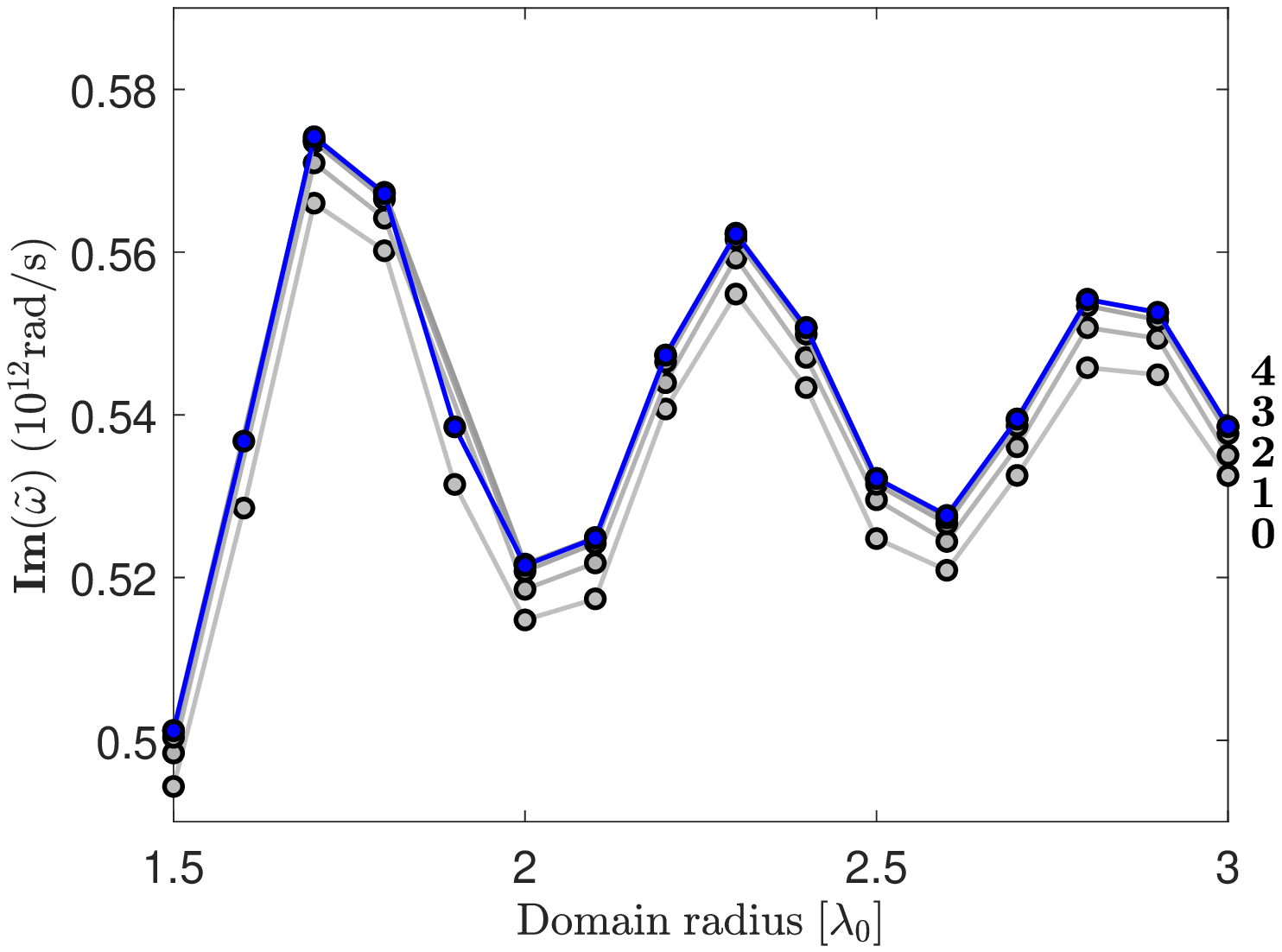}}
  \caption{Numerical oscillations in the real \textbf{(a)} and the imaginary \textbf{(b)} parts of the eigenfrequency when varying the computational domain size.  The oscillations are shown for the five different discretizations used in the convergence study, starting from one initial coarse mesh to the fourth mesh refinement.}
  \label{figA3}
\end{figure}

In Fig.\ref{figA3}, we show the numerical oscillations separately for the real and imaginary parts of the eigenfrequency.  Again, we observe that the oscillations in the quantity of interest shift with improving discretization, and the oscillations tend to become cleaner. We note that for this study the imaginary part of the eigenfrequency is much more sensitive to the computational domain size than to the discretization.

\begin{figure}
  \centering
    \subfloat[\label{figA4a}]{\includegraphics[width=0.43\columnwidth]{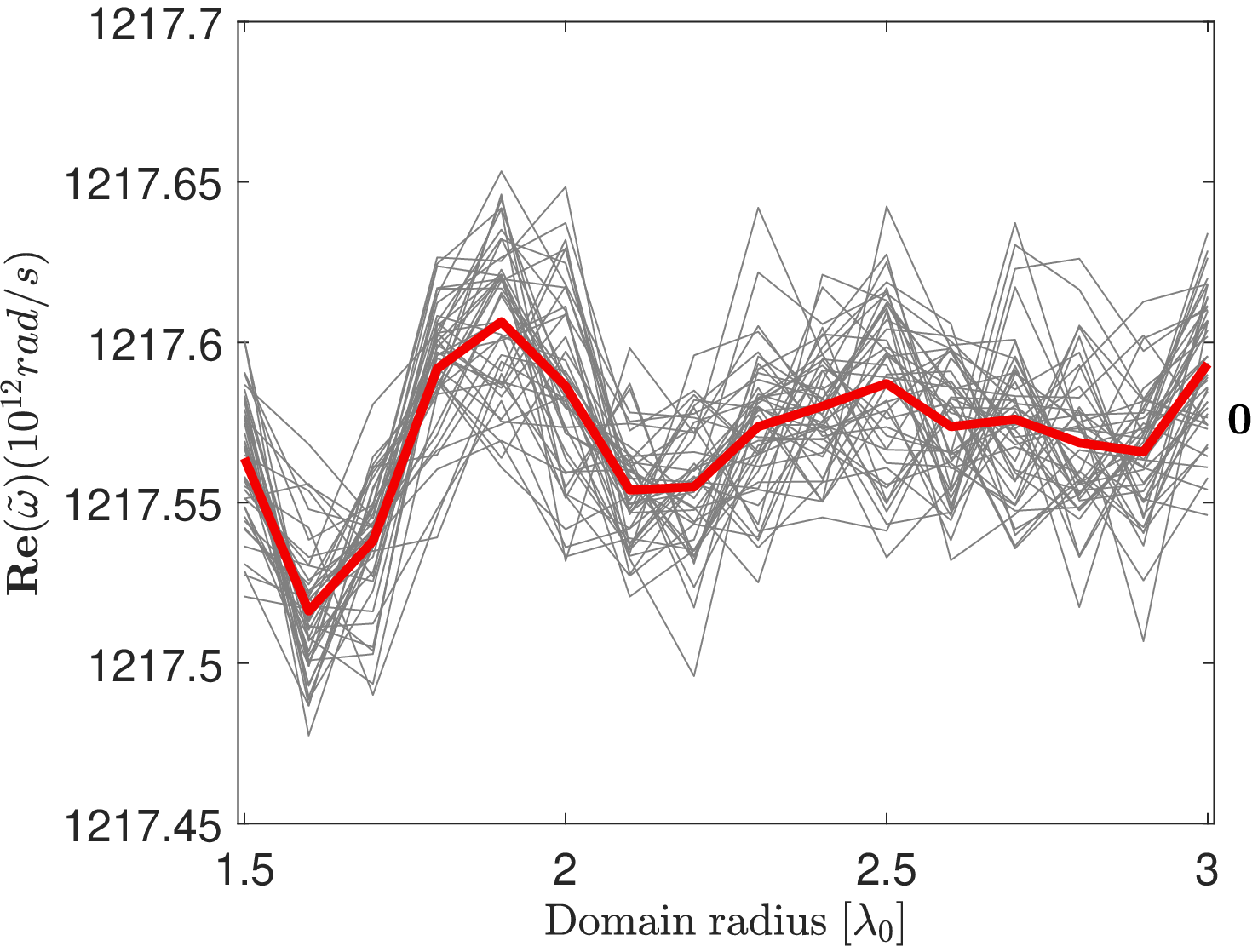}}
    \subfloat[\label{figA4b}]{\includegraphics[width=0.43\columnwidth]{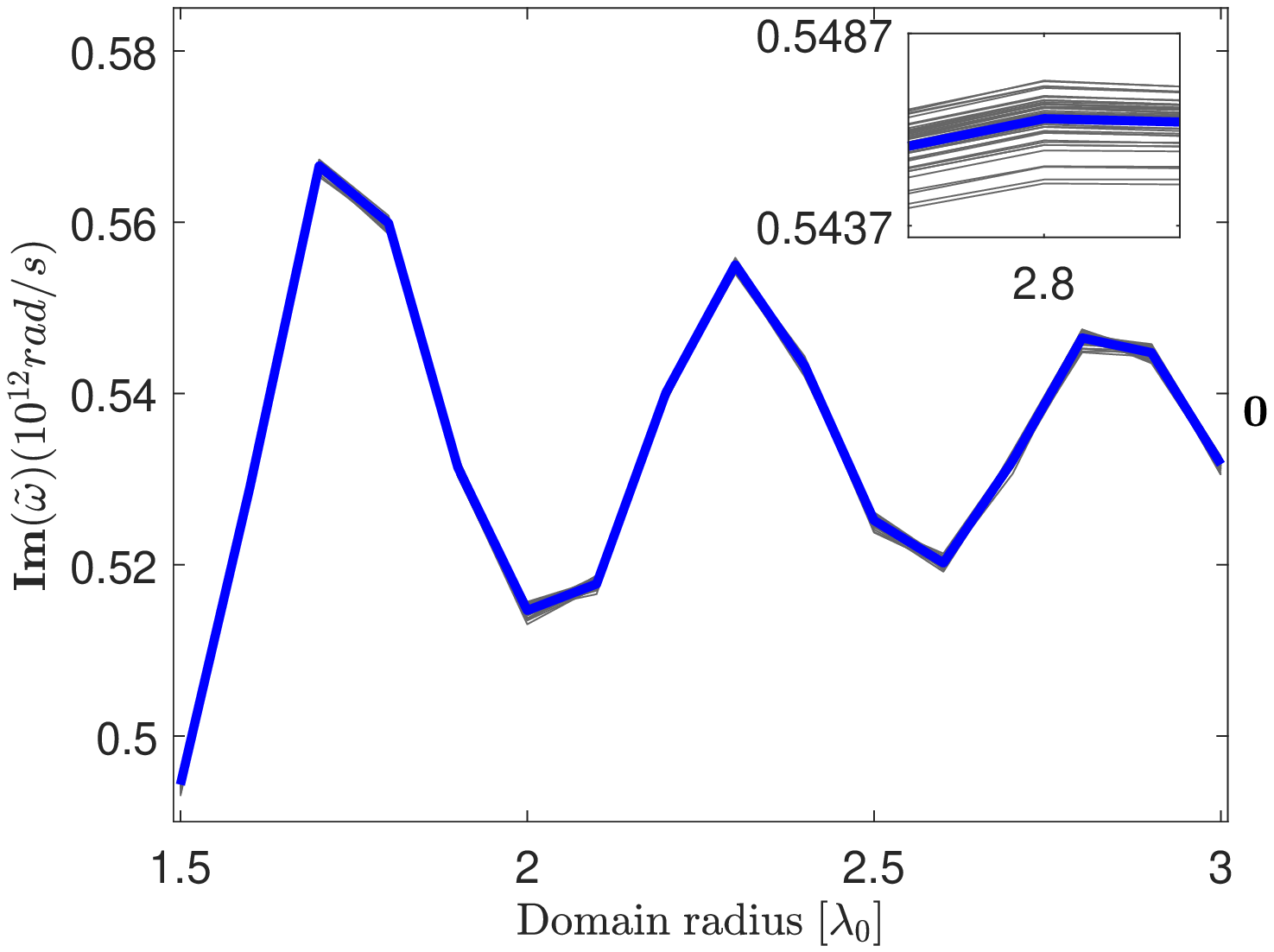}}
  \caption{Sensitivity to the mesh.  Numerical oscillations with domain size in the real \textbf{(a)} and the imaginary \textbf{(b)} parts of the eigenfrequency for the coarsest discretization for different nominally identical meshes.  Thick lines show the averaged quantity for each domain radius, while the gray lines are the results for 44 different nominally identical meshes used for the averaging.  Inset in \textbf{(b)} shows a zoom-in close to the domain radius $2.8\lambda_0$.}
  \label{figA4}
\end{figure}

The noise in the observed oscillations for the coarser meshes can be reduced by averaging over different nominally identical meshes, as illustrated in Fig.\ref{figA4}, which shows the average of the eigenfrequency over 44 different, but nominally identical realizations of the coarsest mesh.  The gray lines are from the individual calculations, which can noticeably vary even for minimal differences in the discretization.  To improve the accuracy of the convergence studies described in the following section, we used this kind of average over sets of nominally identical meshes, although we found no significant difference in the final result.

\subsubsection{Estimating physical quantities}

\begin{figure}
  \centering
    \subfloat{\includegraphics[width=0.5\columnwidth]{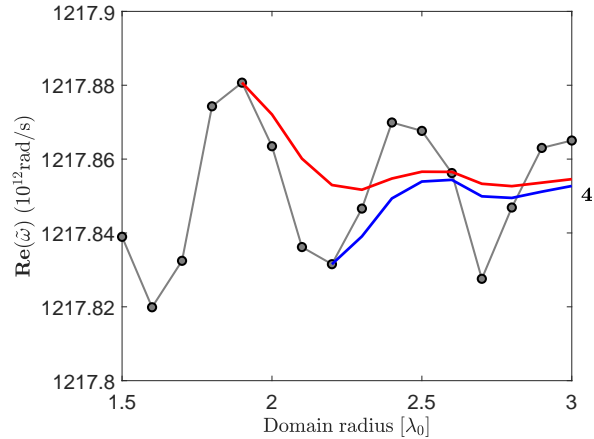}}
  \caption{The running average scheme used to average out the numerical oscillations occuring with varying domain size.}
  \label{figA5}
\end{figure}

To approximate the limiting value from the numerical oscillations we use a running average approach, calculating $\bar{f}(N) = \sum_i^N f_i/N $ starting from a specific point in the oscillations and gradually including more and more points in the averaging.  This running average also oscillates, and we thus estimate the true value by taking the average of the maximum and minimum of the final oscillations.  To reduce error, we take two running averages using this approach, one starting from a minimum of the oscillations, and one starting from a maximum, as illustrated in Fig.\ref{figA5}.  These two estimates are then averaged, and the error is taken to be the half-distance between the individual running average estimates.  Because of the limited number of oscillations and the relatively small size of the oscillations in these running averages, we simply use the two estimates including the greatest number of points in order to estimate the upper and lower limits to the average value.  In this way, we get a single estimate of the quantity of interest for each discretization, which we can use to test convergence with respect to the mesh element size.

\begin{figure}
  \centering
    \subfloat[\label{figA6a}]{\includegraphics[width=0.45\columnwidth]{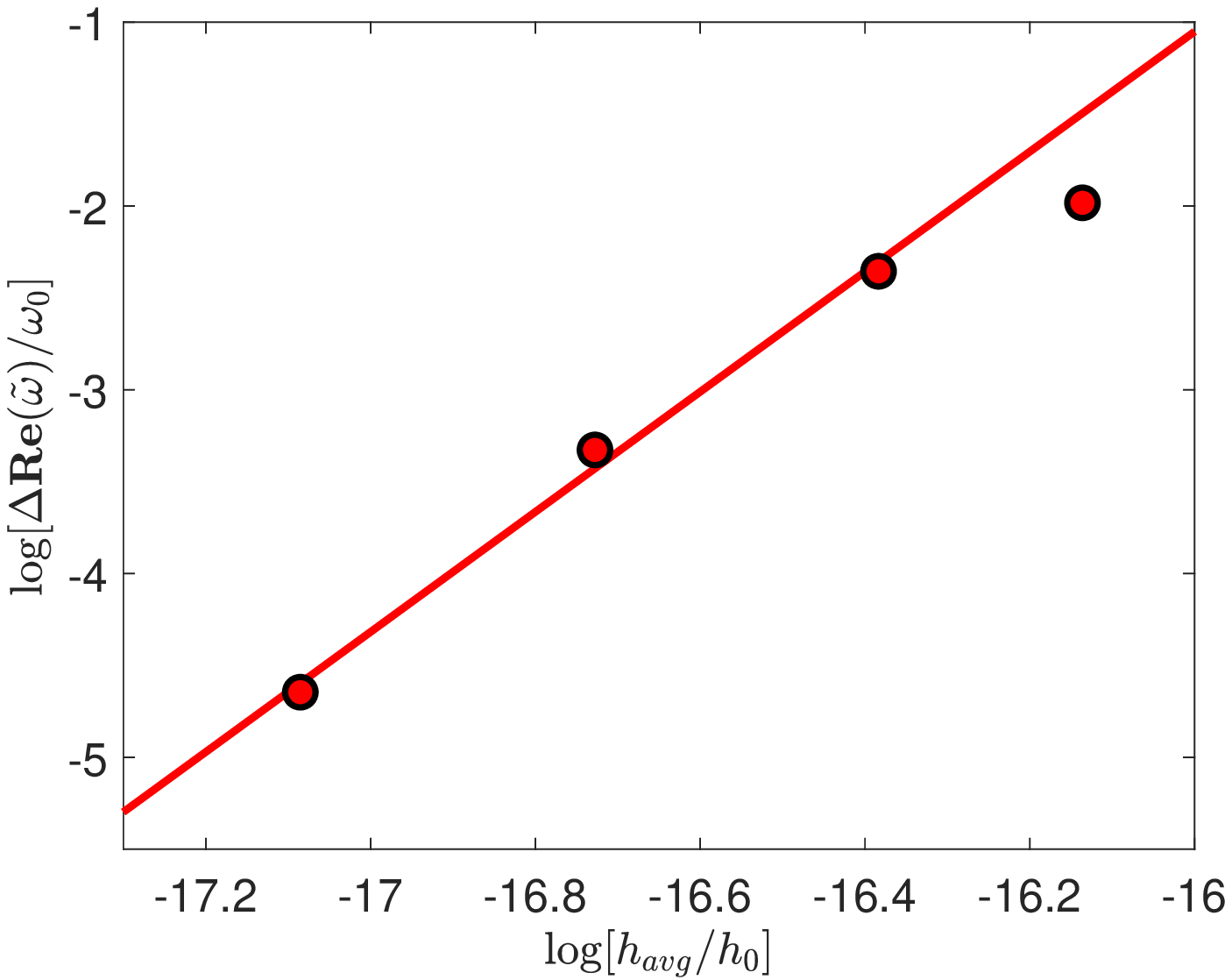}}
    \subfloat[\label{figA6b}]{\includegraphics[width=0.45\columnwidth]{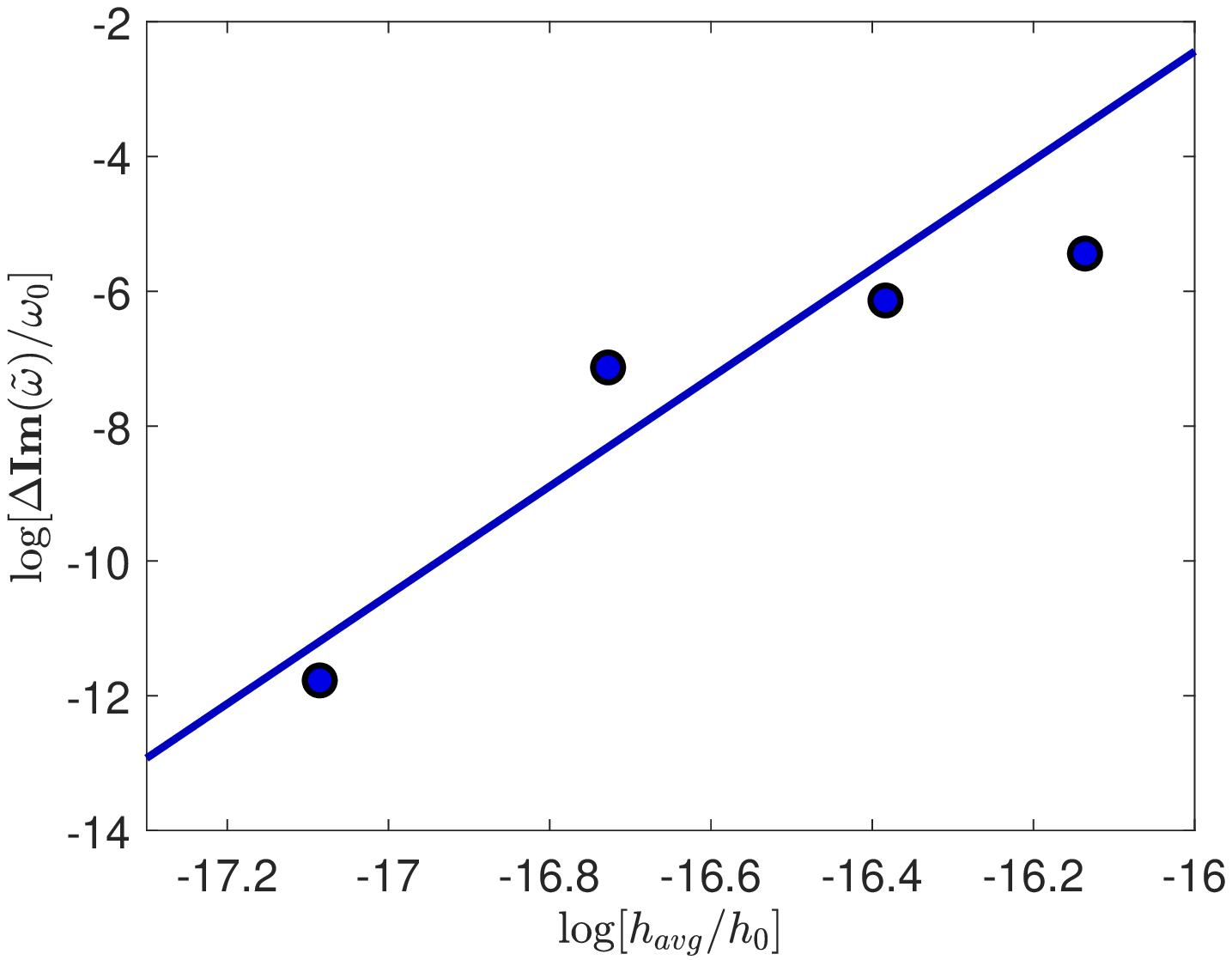}}
  \caption{Mesh convergence study for the real \textbf{(a)} and the imaginary \textbf{(b)} parts of the eigenfrequency.  The horizontal axis shows the logarithm of the average element size, while the vertical axis shows the logarithm of the difference in value between two consecutive discretizations.  The linear fit is between the four finest discretizations; here the first three points in the plots.  The quantities in the logarithms have been made dimensionless through scaling with $h_0=1 \,\mathrm{nm}$ and $\omega_0= 10^{12}\mathrm{rad \, s^{-1}}$.}
  \label{figA6}
\end{figure}

In order to investigate the convergence in a quantitative way, we assume polynomial convergence of the form $f(h) = f_0 + \mathcal{E}_0 h^\alpha$, where $f(h)$ is the calculated value of the quantity of interest, $f_0$ is the true value, and the term $\mathcal{E}_0 h^\alpha$ represents the error at the discretization $h$.  In order to determine the parameters $\mathcal{E}_0$ and $\alpha$, we form the differences $f(h)-f(xh) = \mathcal{E}_0 h^\alpha [1-x^\alpha]$ for meshes of a chosen parameter $h$ differing by some factor $x$. If the discretizations are related by constant ratios $x$, we can now plot these parameters on a double log scale to get a simple linear relationship and thereby determine the unknown constants \cite{Kristensen2020}. As shown in Table \ref{Table1}, the last four discretizations vary systematically, differing by ratios very close to $0.70$.  The linear relationship is expected to hold better at the limit of finer discretizations, and therefore, we use the finer four meshes for the linear fit. This gives rise to the first three points in Figs \ref{figA6a} and \ref{figA6b} which show the mesh convergence study for the real and imaginary parts of the complex eigenfrequency. Using the running average approach, we get the limiting estimates of the complex eigenfrequency $\mathbf{Re}(\tilde{\omega}_i)$ and $\mathbf{Im}(\tilde{\omega}_i)$ for each mesh discretization $h_\text{avg}$, and by plotting the logarithm of the difference of consecutive points against the logarithm of $h_\text{avg}$, we find an approximately linear relationship, which we fit using a least squares method. Using this linear fit, we can finally approximate the true solution $f_0$ along with an associated error from the result with the finest discretization as $\sim\max \left\{ |f_0 - [ f(h_\text{min}) \pm \delta f(h_\text{min})] | \right\}$.  In order to account for the uncertainty in determining the center of the oscillations, we include an additional error as the oscillation half-amplitude of the finest discretization.  In this way, we get a conservative estimate of the uncertainty.

\end{document}